\documentclass[12pt]{iopart}
\usepackage[slovene,english]{babel} 
\usepackage{dsfont}
\usepackage{mathrsfs}

\usepackage{graphicx,bbm}
\usepackage{mathrsfs}
\usepackage{subfigure}
\usepackage{hyperref}
\usepackage{color}
\usepackage[export]{adjustbox}

\newcommand{\mc}[1]{\mathcal{#1}}
\newcommand{\mch}[1]{\hat{\mathcal{#1}}}

\newcommand{\msh}[1]{\hat{\mathscr{#1}}}

\newcommand{\ds}[1]{\mathds{#1}}
\newcommand{\ul}[1]{\underline{#1}}

\newcommand{\trb}[1]{ {\rm tr}\left( #1\right) }
\newcommand{\bra}[1]{ \left \langle #1\right | }
\newcommand{\brao}[1]{ \left ( #1\right | }
\newcommand{\ket}[1]{ \left | #1\right \rangle}

\newcommand{\braoket}[2]{ \left ( #1\middle| #2\right \rangle}

\newcommand{\dd}{ {\rm d} }
\newcommand{\ddt}{\frac{\dd}{\dd t} }
\newcommand{\ii}{ {\rm i} }

\newcommand{\mct}{\mathcal{T}(\mathcal{H})}
\newcommand{\mcb}{\mathcal{B}(\mathcal{H})}

\newcommand{\ave}[1]{{\langle #1\rangle}}

\newcommand{\Aref}[1]{\ref{#1}}

\usepackage[all]{xy}

\begin{document}

\title{Closed hierarchy of correlations in Markovian open quantum systems}

\author{Bojan \v{Z}unkovi\v{c}}
\address{Departamento de F\' isica, Facultad de Ciencias F\' isicas y Matem\' aticas, Universidad de Chile, Casilla 487-3, Santiago Chile}

\date{\today}

\begin{abstract}
We study the Lindblad master equation in the space of operators and provide simple criteria for closeness of the hierarchy of equations for correlations. We separately consider the time evolution of closed and open systems and show that open systems satisfying the {\it closeness conditions} are not necessary of Gaussian type.  In addition, we show that dissipation can induce the closeness of the hierarchy of correlations in interacting quantum systems. As an example we study an interacting optomechanical model, the Fermi-Hubbard model, and the Rabi model, all coupled to a fine-tuned Markovian environment and obtain exact analytic expressions for the time evolution of two-point correlations.
\end{abstract}

\maketitle
\section{Introduction}
A complete description of nonequilibrium steady states is one of the main aims of the nonequilibrium statistical mechanics. In this regard exactly solvable models are of great importance. Their value lies in opportunity to compare numerical simulations and ability to recognize fundamental physical laws. In the classical setting steady states have been numerically and analytically investigated in stochastic lattice gas models \cite{KLS84} and in exactly solvable one-dimensional exclusion processes \cite{DDM92,DEH+93,DJL+93,Der07}, where the steady state is framed in a matrix product form \cite{BE07}. This ansatz has proven useful also in quantum setting, where it has been utilized to construct the nonequilibrium steady state of the boundary driven XXZ spin 1/2 chain \cite{Pro11,Pro11a, KPS13}. Furthermore, the existing classes of Markovian many-body open quantum system models that can be exactly solved (by this we mean to analytically obtain the steady state) are strongly related to integrable closed quantum models, namely the quasi-free bosonic or fermionic models \cite{Pro08,PS10,Kos09} and Yang-Baxter integrable models \cite{IZ13}.  In addition, there are some solutions that do not evidently rely on the underlying integrability structure, e.g., the quantum symmetric exclusion process \cite{TWV12}, the XX spin 1/2 chain with dephasing \cite{Zni10}, and the non-interacting harmonic oscillator lattices in arbitrary dimensions with dephasing \cite{AMT+12}. Exact long-time behavior has been found also in time-dependent driven quadratic systems \cite{PI11,ZZP11}. Whereas, in  the mentioned cases the steady state is obtained as a fixed point of a non-unitary flow, generated by a Liouvillian of Lindblad form, a different approach to nonequilibrium statistical mechanics  applicable to infinitely extended systems has been initiated by Ruelle \cite{Rue69,Rue00, Rue01} and has been actively developed since \cite{aji-Bussei}. A drawback of this approach is that the results are mainly limited to quasi-free systems. Thus, exact analytic results for steady-states of {\it interacting} many-body models out of equilibrium are still scarce, and for {\it non-integrable} systems even scarcer.

In this paper we provide simple criteria (based on quantization in the space of operators) for a closed hierarchy of equations of motion for correlations of Markovian open quantum systems. The closeness property enables efficient calculation of the low-order steady-state correlations, although a complete description of the nonequilibrium steady state may still be difficult to assess. Interestingly, the closeness of correlation hierarchy can be exploited to {\it analytically} obtain exact steady-state correlations even in {\it non-integrable interacting} systems. We demonstrate this by calculating the two-point correlations in three paradigmatic interacting models, namely an interacting optomechanical model, the Fermi-Hubbard model, and the Rabi model. Solutions of these and similar out-of-equilibrium models can be important beyond the nonequilibrium setting, since they provide a new insight into the structure of interacting models.

\section{Preliminaries}
Before we proceed let us shortly introduce the method of quantization in the space of operators upon which the closeness criteria are based. We want to study the  Lindblad master equation 
\begin{eqnarray}
\label{eq:lind}
\ddt \rho=-\ii[\rho,H]+\sum_k L_k\rho L^\dag_k-\frac{1}{2}\{L^\dag_k L_k,\rho\},
\end{eqnarray}
where $\{A,B\}=AB+BA$ represents the anti-commutator, $[A,B]=AB-BA$ the commutator, $H$ is the Hamiltonian of the system, and $L_k$ are the Lindblad operators. The first part on the right hand side of \eref{eq:lind} generates the unitary evolution and the rest is responsible for dissipation. Our goal is to efficiently rewrite the Lindblad master equation \eref{eq:lind} in the Liouville space by lifting the methods of second quantization to the space of operators. Since we want to describe fermions and bosons with the same formalism we have to introduce the space of trace class operators $\mct$ and the space of bounded operators $\mcb$ acting on the many-body Hilbert space $\mc{H}$. For each $A\in \mcb$ we can define a functional $\brao{A}$ acting on $\ket{\rho}\in\mct$ as 
\begin{eqnarray}
\label{eq:def funct}
\braoket{A}{\rho}=\trb{A^\dag\rho}.
\end{eqnarray} 
The functionals $\brao{A}$ form a dual space $\mct^*$ \cite{AL06}. We use the bra-ket notation and indicate the element of $\mct$ with $\ket{\rho}$ and the element of $\mct^*$ with $\brao{A}$. The super-operators acting on the spaces $\mct$ from the left and $\mct^*$ from the right shall be marked with a hat $\hat{\bullet}$. When expressing the action of the Liouvillian and dissipators on the ket vectors $\ket{\rho}$ we shall omit the ket and write simply $\rho$, as in equation \eref{eq:lind}.  In order to rewrite the Liouvillian, which determines the right hand side of \eref{eq:lind}, in an efficient and clear way we introduce two fundamental super-operator maps, namely the left $\msh{L}$ and the right $\msh{R}$ multiplication maps defined as
\begin{eqnarray}
&\msh{L},\msh{R}: \mcb \times \mct \rightarrow \mct, \\ \nonumber
&\msh{L}(A)\ket{\rho}=\ket{A\rho},\quad \msh{R}(A)\ket{\rho}=\ket{\rho A},\quad A\in\mcb.
\end{eqnarray}
Their action on the dual space can be deduced from the definition of the functionals (\ref{eq:def funct}). All bosonic (fermionic) Liouvillians of the form \eref{eq:lind} can be rewritten with the aid of left $\msh{L}$ and right $\msh{R}$ multiplication maps of creation $a_j^\dag$ ($c_j^\dag$) and annihilation $a_j$ ($c_j$) operators satisfying the canonical (anti-)commutation relations; $[a_j,a_k^\dag]=\delta_{jk},$ $[a_j,a_k]=[a_j^\dag,a_k^\dag]=0$ ; ($\{c_j,c_k^\dag\}=\delta_{jk},$ $\{c_j,c_k\}=\{c_j^\dag,c_k^\dag\}=0$). We use these sets of left and right multiplication maps to define creation and annihilation maps satisfying almost canonical commutation relations (almost-CCR) $[\hat{b}_{j},\hat{b}_{k}^+]=\delta_{jk},$ $[\hat{b}_{j},\hat{b}_{k}]=[\hat{b}_{j}^+,\hat{b}_{k}^+]=0$, defined via left and right multiplication maps as
\begin{eqnarray}
\label{eq:mb}
\hat{b}_{j}^+&=\msh{L}(a_j^\dag)-\msh{R}(a_j^\dag),\quad\hat{b}_{j}=\msh{L}(a_j),\\ \nonumber
\hat{b}_{N+j}^+&=-\msh{L}(a_j)+\msh{R}(a_j),\quad \hat{b}_{N+j}=\msh{R}(a_j^\dag),
\end{eqnarray}
and almost canonical anti-commutation relations (almost-CAR) $\{\hat{f}_{j},\hat{f}_{k}^+\}=\delta_{jk},$ $\{\hat{f}_{j},\hat{f}_{k}\}=\{\hat{f}_{j}^+,\hat{f}_{k}^+\}=0$ in the fermionic case, defined as
\begin{eqnarray}
\label{eq:mf}
\hat{f}_{j}^+&=&\frac{1}{\sqrt{2}}(-\msh{L}(c_j^\dag)+\msh{R}(c_j^\dag))\mch{P},
\quad \hat{f}_{j}=\frac{1}{\sqrt{2}}(\msh{L}(c_j)+\msh{R}(c_j))\mch{P},\\ \nonumber
\hat{f}_{N+j}^+&=&\frac{1}{\sqrt{2}}(\msh{L}(c_j)-\msh{R}(c_j))\mch{P},
\quad 
 \hat{f}_{N+j}=\frac{1}{\sqrt{2}}(-\msh{L}(c_j^\dag)-\msh{R}(c_j^\dag))\mch{P},
\end{eqnarray}
where $\mch{P}=\msh{L}(P)$ is the parity super-operator, $P=\prod_{j=1}^N(c_jc_j^\dag-c_j^\dag c_j)$ is the parity operator, and $N$ is the number of bosonic or fermionic modes. The word \textit{almost} indicates that the plus (+) in $\hat{b}^+_{j}$ or $\hat{f}^+_{j}$ indicates the bosonic or fermionic creation maps and \textit{not} the adjoint maps (with respect to the inner product \eref{eq:def funct}) of $\hat{b}_j$ or $\hat{f}_j$, respectively. The equations (\ref{eq:mb}) and (\ref{eq:mf}) can be inverted and we can rewrite all Liouvillians of the Lindblad form in terms of creation and annihilation maps satisfying almost-CCR (almost-CAR) for bosons (fermions).

In the following we shall use the super-operators in \eref{eq:mb} and \eref{eq:mf} to determine the {\it closeness conditions} for the hierarchy of equations for the correlation tensors in the bosonic, fermionic and mixed case. In each case we separately consider the generators of unitary and non-unitary dynamics. We show that generators of the dissipative dynamics satisfying the closeness condition are not restricted to quadratic models, as in the unitary case. In the fermionic and bosonic case we obtain general conditions for the closeness of the hierarchy for quadratic noise, whereas in the mixed case we show a simple example of quadratic noise with a closed hierarchy of equations. Finally, at the end of each section, we study three examples of interacting Hamiltonians, which by themselves do not exhibit a closed hierarchy, but in the presence of fine-tuned dissipation satisfy the closeness conditions. In particular, we find  exact expressions for the time evolution of low-order correlations. 

\section{Bosonic Liouvillians}
We begin with the bosonic case, which seems simpler due to canonical commutation relations. 
\subsection{Bosonic closeness condition}
As discussed in the previous section we can always rewrite the bosonic Liouvillian $\mch{B}$ in terms of bosonic creation and annihilation maps
\begin{eqnarray}
\label{eq:liouv}
\mch{B}&=\sum_{m,n}\mch{B}^{(m,n)},\\ \nonumber
\mch{B}^{(m,n)}&=\sum_{j_1\ldots j_m,k_1\ldots k_n}B^{(m,n)}_{j_1\ldots j_m,k_1\ldots k_n}\hat{b}_{j_1}^+\ldots\hat{b}_{j_m}^+\hat{b}_{k_1}\ldots\hat{b}_{k_n}.
\end{eqnarray}
The non-negative integers $m$ and $n$ denote the number of creation and annihilation maps in the super-operators $\mch{B}^{(m,n)}$ and the lengths of the vectors $\ul{j}$ and $\ul{k}$, respectively. An important property of the creation maps $\hat{b}_{j}^+$ defined in \eref{eq:mb} is that they left-annihilate the identity operator, namely 
\begin{eqnarray}
\label{eq:vac left}
\brao{\mathds{1}}\hat{b}^+_j=0. 
\end{eqnarray}
The trace preservation property $\brao{\mathds{1}}\mch{B}=0$ and the  condition (\ref{eq:vac left}) imply that in the sum (\ref{eq:liouv}) we always have $m>0$. 
Let us now define a symmetric $p-$point correlation tensor (correlator) as
\begin{eqnarray}
\label{eq:ref z}
Z^{(p)}_{j_1,j_2,\ldots,j_p}(t)=\brao{1}\hat{b}_{j_1}\hat{b}_{j_2}\ldots\hat{b}_{j_p}\ket{\rho(t)}.
\end{eqnarray}
The time evolution of correlator ${\bf Z}^{(p)}$ is determined by the equation
\begin{eqnarray}
\label{diff_boson}
\ddt Z^{(p)}_{j_1,j_2,\ldots,j_p}(t)&=\brao{1}\hat{b}_{j_1}\hat{b}_{j_2}\ldots\hat{b}_{j_p}\mch{B}\ket{\rho(t)}.
\end{eqnarray}
Using the almost-CCR and the property (\ref{eq:vac left}) we observe that the expression $\brao{1}\hat{b}_{j_1}\hat{b}_{j_2}\ldots\hat{b}_{j_p}\mch{B}\ket{\rho(t)}$ is a function of correlators ${\bf Z}^{(p+M_{\min})},{\bf Z}^{(p+M_{\min}+1)},\ldots {\bf Z}^{(p+M_{\max})}$ only, with $M_{\min}=\min(n-m)$ and $M_{\max}=\max(n-m)$, where the minimum and maximum are taken over all paris $(m,n)$ for which $\mch{B}^{(m,n)}$  in equation \eref{eq:liouv} is not zero. The property \eref{eq:vac left} implies that all correlators ${\bf Z}^{(p-m+n)}$ with $p-m+n<0$ vanish. Therefore, the time evolution for the correlation tensors is closed iff $M_{\max}\leq0$ ({\it bosonic closeness condition} for the hierarchy of equations). For quadratic Hamiltonians and linear Lindblad operators we have $m+n=2$ and since $m>0$ the closeness condition is always satisfied. This has been exploited to find the properties of driven noninteracting bosonic systems \cite{PS10}  and simple harmonic chains coupled to heat baths \cite{ZP12}. Moreover, as we shall shortly show, the closeness condition for bosons can be satisfied also in a more general setting, e.g., harmonic chains in arbitrary dimension and in the presence of dephasing \cite{AMT+12}. 

\subsection{Unitary evolution}
In this section we study as an example an interacting fourth order Hamiltonian of the form $H=\sum_{i,j,k,l=1}^NH_{ijkl}a_i^\dag a_j^\dag a_ka_l$. Due to  Hermicity of the Hamiltonian we have $H_{i,j,k,l}=\bar{H}_{l,k,j,i}$, where the bar $\bar{\bullet}$ denotes complex conjugation. By using \eref{eq:mb} we write the adjoint map of the Hamiltonian $\hat{\rm{ad}}_{H}:=[H,\bullet]$ in the form (\ref{eq:liouv})
\begin{eqnarray}
\label{eq:int h}
\mch{B}_{\rm int}&=&-\ii\,\hat{ad}_{H}=\mch{B}_{\rm int}^{(1,3)}+\sum_{m\geq n}\mch{B}_{\rm int}^{(m,n)}, \\ \nonumber
\mch{B}_{\rm int}^{(1,3)}&=&-\ii\sum_{i,j,k,l}\left(H_{ijkl}\, \hat{b}_{i}^+ \hat{b}_{N+j} \hat{b}_{k} \hat{b}_{l} + \,\hat{b}_{N+k}^+ \hat{b}_{N+i} \hat{b}_{N+j} \hat{b}_{l}\right).
\end{eqnarray}
The only term violating the bosonic closeness condition is $\mch{B}^{(1,3)}$. By using the symmetry of the tensor  ${\bf H}$ with respect to permutation of the first and the last two indices we find that $\mch{B}^{(1,3)}=0$ implies $H=0$. Not surprisingly, similar symmetry arguments can be applied for a higher order Hamiltonians showing that a hierarchy of equations for correlations cannot be closed for any non-quadratic (or linear) bosonic Hamiltonian (see the \Aref{closeness_hamilton}). 

\subsection{Dissipation}
In contrast to the unitary case, dissipators can contain a product of more than two creation and annihilation maps and nevertheless exhibit a closed hierarchy of equations for correlations. To illustrate this we consider an example of a purely dissipative evolution determined by a Lindblad operator $L=\sum_{j,k}A_{jk}d_j^\dag d_k$. The corresponding dissipative Liouvillian may be written as
\begin{eqnarray}
\label{eq:noise}
\mch{B}_{\rm dis}&=&\mch{B}_{\rm dis}^{(1,3)}+\sum_{m\geq n}\mch{B}_{\rm dis}^{(m,n)}, \\ \nonumber
\mch{B}_{\rm dis}^{(1,3)}&=&\frac{1}{2}\sum_{i,j,k,l}\Big(\left(A_{il} \bar{A}_{kj}-A_{j,l} \bar{A}_{k,i}\right) \hat{b}_{i}^\dag \hat{b}_{N+j} \hat{b}_{k} \hat{b}_{l}\\ \nonumber
 &&~~~~~~~~~~+ \left(A_{ki} \bar{A}_{lj}-A_{kl} \bar{A}_{ij}\right) \hat{b}_{N+i}^\dag \hat{b}_{N+j} \hat{b}_{N+k} \hat{b}_{l}\Big).
\end{eqnarray}
The {\it closeness condition} $M_{\max}=0$ implies 
\begin{eqnarray}
\label{eq:2dis}
A_{il} \bar{A}_{kj}-A_{j,l} \bar{A}_{k,i}+A_{ik} \bar{A}_{lj}-A_{j,k} \bar{A}_{l,i}=0,
\end{eqnarray}
and is satisfied by quadratic Hermitian Lindblad operators.  In fact, Hermitian ${\bf A}$ is the sole solution of equation (\ref{eq:2dis}), apart from an unimportant phase factor, showing that in the considered case quadratic Hermitian Lindblad operators are necessary and sufficient condition for closed hierarchy of correlation equations. 

In in general we can discuss the closeness for a dissipator of the form
\begin{eqnarray}
\label{eq:general_D}
&\mch{D}=\sum_{\mu,\nu,\mu',\nu'}\sum_{\ul{j},\ul{k},\ul{j}',\ul{k}'}G^{(\mu,\nu;\mu',\nu')}_{\ul{j},\ul{k};\ul{j}',\ul{k}'}\mch{D}(L_{\ul{j},\ul{k}}, L_{\ul{j}',\ul{k}'}),\\ \nonumber
&\mch{D}(L_{\ul{j},\ul{k}}, L_{\ul{j}',\ul{k}'})\rho=L_{\ul{j},\ul{k}}\rho L_{\ul{j}',\ul{k}'}^\dag-\frac{1}{2}\{L_{\ul{j}',\ul{k}'}^{\dag} L_{\ul{j},\ul{k}},\rho\},
\end{eqnarray}
where $\bf{G}$ is a positive Hermitian rate matrix, $\mu,\,\nu,\,\mu',\,\nu'$ determine the lengths of vectors  $\ul{j},\,\ul{k},\,\ul{j}',\,\ul{k}'$, respectively, (e.g. $\ul{j}=(j_1,j_2\ldots,j_\mu)$). In the bosonic case we may take the following coupling operators
\begin{eqnarray}
L_{\ul{j},\ul{k}}=a^\dag_{j_1}a^\dag_{j_2}\ldots a^\dag_{j_\mu} a_{k_1}a_{k_2}\ldots a_{k_\nu}.
\end{eqnarray}
In the \Aref{closeness_diss} we show how the closeness condition imposes additional symmetry constraints to the rate matrix $\bf{G}$. We find that the simplest additional requirements are satisfied by rate matrix with the following symmetry
\begin{eqnarray}
\label{eq:boson_quad_bath}
G^{(\mu,\nu;\mu',\nu')}_{\ul{j},\ul{k};\ul{j}',\ul{k'}}-\bar{G}^{(\nu,\mu;\nu',\mu')}_{\ul{k},\ul{j};\ul{k}',\ul{j}'}=0.
\end{eqnarray}
In case of quadratic noise, i.e., restricting the set of coupling operators to $L_{j,k}\in\{a_ja_k, a_j^\dag a_k^\dag, a_j^\dag a_k\}$, the rate matrix $\bf{G}$ with the symmetries \eref{eq:boson_quad_bath} has a closed hierarchy of equations. It remains an open question if \eref{eq:boson_quad_bath} is the only solution of the general conditions presented in the \Aref{closeness_diss}.

\subsection{Exact solution of a dissipative interacting Boson model}
In this section we shall combine the results of the previous two sections and devote our attention to a combined unitary and dissipative evolution. The idea is to show that an interacting bosonic Hamiltonian can satisfy the closeness condition if coupled to fine-tuned Lindblad dissipators. For this purpose we study the simplest interacting bosonic Hamiltonian of the form
\begin{eqnarray}
\label{optm_H}
H_{\rm VR}=\omega \,a_1^\dag a_1+\Omega \,a_2^\dag a_2 + g \,a_1^\dag a_1(a_2+a_2^\dag),
\end{eqnarray}
which represents the interaction between one vibrational and one radiation mode, and successfully describes most of the observed optomechanical phenomena \cite{AKM13}. The action of the adjoint map $\hat{\rm ad}_{H_{\rm VR}}$ can be expressed in terms of the bosonic creation and annihilation super-operators \eref{eq:mb} as
\begin{eqnarray}
\label{ad_optm}
\hat{\rm ad}_{H_{\rm VR}}=&\omega(\hat{b}_1^+\hat{b}_1-\hat{b}_3^+\hat{b}_3)+\Omega(\hat{b}_2^+\hat{b}_2-\hat{b}_4^+\hat{b}_4)\\ \nonumber
&+g\Big(\hat{b}^+_1 \hat{b}_2 \hat{b}_1 + \hat{b}^+_1 \hat{b}_4 \hat{b}_1 + \hat{b}^+_2 \hat{b}_3 \hat{b}_1 - \hat{b}^+_3 \hat{b}_3 \hat{b}_2  \\ \nonumber
&~~~~~- \hat{b}^+_3 \hat{b}_4 \hat{b}_3 - \hat{b}^+_4 \hat{b}_3 \hat{b}_1 +  \hat{b}^+_2 \hat{b}^+_1 \hat{b}_1 - \hat{b}^+_4 \hat{b}^+_3 \hat{b}_3\Big).
\end{eqnarray} 
Evidently, all but the last two interacting terms do {\it not} preserve the hierarchy of correlations. We wish to cancel them by adding  dissipation of the form \eref{eq:general_D} with the coupling operator basis $L_{\ul{j},\ul{k}}\in \{ a_1^\dag,a_2^\dag,a_1,a_2,a^\dag_1a_2^\dag,a^\dag_2a_1,a^\dag_1a_1,a_2a_1,a^\dag_1a_2 \}$. It can be shown that the rate matrix
\begin{eqnarray}
\label{g_boson}
{\bf G}=\left(\begin{array}{ccccccccc}
\Gamma_1^+&0&0&0&-2 \ii g&0&0&0&0\\ 
0&\Gamma_2^+&0&0&0&0&0&0&0\\ 
0&0&\Gamma_1&0&0&-2 \ii g&0&0&0\\ 
0&0&0&\Gamma_2&0&0&-2 \ii g&0&0\\ 
2\ii g&0&0&0&\Gamma_{1,2}&0&0&0&0\\
0&0&2\ii g &0&0&\Gamma_{2,1}&0&0&0\\ 
0&0&0&2\ii g &0&0&\Gamma_{1,1} &0&0\\
0&0&0&0&0&0&0&\Gamma_{1,2}&0\\
0&0&0&0&0&0&0&0&\Gamma_{2,1}
\end{array}\right)
\end{eqnarray}
exactly cancels the unwanted terms in the adjoint map \eref{ad_optm}. For positive rates $\Gamma_1^+,\Gamma_2^+,\Gamma_{1,2}\Gamma_{2,1},\Gamma_{1,1}>0$, and $4g^2<\min(\Gamma_1^+\Gamma_{1,2},\,\Gamma_1\Gamma_{2,1},\,\Gamma_2\Gamma_{1,1})$ the rate matrix \eref{g_boson} is positive and defines a valid Lindblad dissipator $\mch{D}_{\rm VR}$, which with the adjoint map \eref{ad_optm} determines the complete Liouvillian $\mch{B}_{\rm VR}=-\ii\,\hat{\rm ad}_{H_{\rm VR}}+\mch{D}_{\rm VR}$. With the aid of creation and annihilation super-operators \eref{eq:mb} we obtain

\begin{eqnarray}
\label{closed_boson_int}
\mch{B}_{\rm VR}=&-\ii \omega(\hat{b}_1^+\hat{b}_1-\hat{b}_3^+\hat{b}_3)-\ii \Omega(\hat{b}_2^+\hat{b}_2-\hat{b}_4^+\hat{b}_4)\\ \nonumber
&+\,\ii\, g (\hat{b}^+_2 - \hat{b}^+_4 -  2 (\hat{b}^+_2 \hat{b}^+_1 \hat{b}_1 + \hat{b}^+_3 \hat{b}^+_1 \hat{b}_2 - \hat{b}^+_3 \hat{b}^+_1 \hat{b}_4 - 
      \hat{b}^+_3 \hat{b}^+_2 \hat{b}_3 \\ \nonumber 
 &~~~~~+ \hat{b}^+_4 \hat{b}^+_1 \hat{b}_1 -  \hat{b}^+_4 \hat{b}^+_3 \hat{b}_3 +\hat{b}^+_4 \hat{b}^+_3 \hat{b}^+_1- \hat{b}^+_3 \hat{b}^+_2 \hat{b}^+_1 ))\\ \nonumber
&+\frac{1}{2} \Bigg( ( \Gamma_{1,2} + \Gamma_1^+- \Gamma_{2,1} - \Gamma_1 - \Gamma_{1,1}) \left(\hat{b}_1^+ \hat{b}_1+\hat{b}_3^+ \hat{b}_3\right)\\ \nonumber
&~~~~~ + (\Gamma_{1,2} +  \Gamma_2^+- \Gamma_{2,1}-  \Gamma_2 ) \left(\hat{b}_2^+ \hat{b}_2+\hat{b}_4^+ \hat{b}_4\right) \\ \nonumber
&~~~~~+  \Gamma_{1,1} \left( 2  \hat{b}_3^+ \hat{b}_1^+ \hat{b}_3 \hat{b}_1  -\hat{b}_1^+ \hat{b}_1^+ \hat{b}_1 \hat{b}_1 - 
    \hat{b}_3^+ \hat{b}_3^+ \hat{b}_3 \hat{b}_3\right)\Bigg)
\\ \nonumber
&  +    (\Gamma_{1,2} + \Gamma_{2,1})\left(  \hat{b}_3^+ \hat{b}_1^+ \hat{b}_4 \hat{b}_2  +\hat{b}_4^+ \hat{b}_2^+ \hat{b}_3 \hat{b}_1\right) +\Gamma_{1,2}\left( \hat{b}_3^+ \hat{b}_2^+ \hat{b}_3 \hat{b}_2  +   \hat{b}_4^+ \hat{b}_1^+ \hat{b}_4 \hat{b}_1\right) \\ \nonumber
& -\Gamma_{2,1}\left( \hat{b}_2^+ \hat{b}_1^+ \hat{b}_2 \hat{b}_1  +   \hat{b}_4^+ \hat{b}_3^+ \hat{b}_4 \hat{b}_3\right) 
 +   (\Gamma_{1,2}  +\Gamma_1^+) \hat{b}_3^+ \hat{b}_1^+ + (\Gamma_{1,2}  + \Gamma_{2}^+)\hat{b}_4^+ \hat{b}_2^+  \\ \nonumber
&+      \Gamma_{1,2} \left(\hat{b}_3^+ \hat{b}_2^+ \hat{b}_1^+ \hat{b}_2 + \hat{b}_4^+ \hat{b}_2^+ \hat{b}_1^+ \hat{b}_1 + 
         \hat{b}_4^+ \hat{b}_3^+ \hat{b}_1^+ \hat{b}_4 + \hat{b}_4^+ \hat{b}_3^+ \hat{b}_2^+ \hat{b}_3 + 
    \hat{b}_4^+ \hat{b}_3^+ \hat{b}_2^+ \hat{b}_1^+\right).
\end{eqnarray}
Since $m-n>0$ for all terms in the Liouvillian \eref{closed_boson_int}  the bosonic closeness condition is satisfied. This enables us to solve the differential equations \eref{diff_boson} with $\mch{B}_{\rm VR}$ given by \eref{closed_boson_int}. In order to shorten the expressions we further restrict the rates $\Gamma_1^+=\Gamma_2^+=\Gamma_{1,2}=\Gamma_{2,1}=\Gamma^+$ and $\Gamma_1=\Gamma_2=\Gamma_{1,1}=\Gamma$. We assume that at initial time $t=0$ the first and second mode are in a particle vacuum state, and find the following expressions for time evolution of non-vanishing first and second order correlations
\begin{eqnarray}
\ave{a_2}&=&-\frac{2 i g \left(-1+e^{-\frac{1}{2} t \left(-\Gamma ^++\Gamma +2 i
   \Omega \right)}\right)}{-\Gamma ^++\Gamma +2 i \Omega },\\ \nonumber
 \ave{a_2a_2}&=&-\frac{4 g^2 \left(-1+e^{-\frac{1}{2} t \left(-\Gamma ^++\Gamma +2 i
   \Omega \right)}\right)^2}{\left(-\Gamma ^++\Gamma +2 i \Omega
   \right)^2}.
\end{eqnarray}
Since the remaining expressions for the time evolution of the occupations are too lengthy we write only their steady state values
\begin{eqnarray}\nonumber
   \ave{a_1^\dag a_1}&=&\frac{2 \Gamma ^+ \left(\Gamma ^++\Gamma \right)+8 \Gamma  \left(\Gamma
   -\Gamma ^+\right) g^2}{\left(\Gamma -3 \Gamma ^+\right) \left(\Gamma
   ^++\Gamma \right)}, \\ \nonumber
\ave{a_2^\dag a_2}&=&\frac{2 \Gamma ^+ \left(\Gamma ^++\Gamma \right)+4 \left(\Gamma -\Gamma
   ^+\right) \left(3 \Gamma ^++\Gamma \right) g^2}{\left(\Gamma -3
   \Gamma ^+\right) \left(\Gamma ^++\Gamma \right)}.
\end{eqnarray}
All remaining one and two-point correlations are zero. From the time evolution of the occupations (not shown) we see that the system is stable if $\Gamma>3\Gamma^+$. Notice that the hierarchy of equations for the considered model can be closed also by a different dissipator. In this regard it would be interesting to see how the properties of the systems change by changing the non-unitary part of the evolution.
\section{Fermionic Liouvillians}
In this section we define the fermionic closeness condition for unitary and dissipative dynamics and show by solving the dissipative Fermi-Hubbard model in one, two, and three dimensions that fine-tuned dissipation can induce a closed hierarchy of equations also in the fermionic case.
\subsection{Fermionic closeness conditions}
As in the previous section we decompose the fermionic Liouvillian $\mch{F}$ in terms of creation and annihilation maps \eref{eq:mf}
\begin{eqnarray}
\label{eq:liouv_fermi}
\mch{F}&=&\sum_{m,n}\mch{F}^{(m,n)}\mch{P}^{m+n},\\ \nonumber
\mch{F}^{(m,n)}&=&\sum_{\ul{j},\ul{k}}L^{(m,n)}_{j_1\ldots j_n,k_1\ldots k_m}\hat{f}_{j_1}^+\ldots\hat{f}_{j_m}^+\hat{f}_{k_1}\ldots\hat{f}_{k_n}.
\end{eqnarray}
The fermionic creation maps $\hat{f}_{j}^+$ defined in \eref{eq:mf} left annihilate the identity operator, namely 
\begin{eqnarray}
\label{eq:vac left_f}
\brao{\mathds{1}}\hat{f}^+_j=0,
\end{eqnarray}
which again implies, by using the trace preservation property $\brao{\mathds{1}}\mch{F}=0$, that in the sum (\ref{eq:liouv_fermi}) we always have $m>0$. 
We define the anti-symmetric $p-$point correlation tensors as
\begin{eqnarray}
\label{fermion_Z}
Z^{(p,\alpha)}_{j_1,j_2,\ldots,j_p}(t)=\brao{1}\hat{f}_{j_1}\hat{f}_{j_2}\ldots\hat{f}_{j_p}\mch{P}^\alpha \ket{\rho(t)},
\end{eqnarray}
with $\alpha=0,1$ determining the presence of the super-operator $\mch{P}$. An additional correlation tensor is necessary because of a parity super-operator in the fermionic Liouvillian \eref{eq:liouv_fermi}. The time evolution of correlation tensors ${\bf Z}^{(p,\alpha)}$ is induced by 
\begin{eqnarray}
\label{diff_Z_fermi}
\ddt Z^{(p,\alpha)}_{j_1,j_2,\ldots,j_p}(t)&=\brao{1}\hat{f}_{j_1}\hat{f}_{j_2}\ldots\hat{f}_{j_p}\mch{P}^\alpha\mch{F}\ket{\rho(t)}.
\end{eqnarray}
Due to the phase factor in the Liouvillian and two different types of correlations it is more difficult to find the closeness conditions than in the bosonic case. The structure of the derivative \eref{diff_Z_fermi} is best explained by the following diagrams

\begin{eqnarray}
\label{diagrams_fermi_a}
\!\!\!\!   \xymatrix{
        {\bf Z}^{(p,0)} \ar[drrr]|{\mch{F}^{(m,n)}\mch{P}} \ar[rrr]^{\!\!\!\!\!\!\!\!\!\!\mch{F}^{(m,n)}} & &&{\bf Z}^{(p-m+n,0)} \\
                             &  & & {\bf Z}^{(p-m+n,1)} } \quad
     \xymatrix{
         & &&{\bf Z}^{(p+m-n,0)} \\
	{\bf Z}^{(p,1)} \ar[urrr]|{\mch{F}^{(m,n)}\mch{P}} \ar[rrr]_{\!\!\!\!\!\!\!\!\!\!\mch{F}^{(m,n)}}&  & & {\bf Z}^{(p+m-n,1)} }
	\label{diagrams_fermi_b}
\end{eqnarray}
The diagrams \eref{diagrams_fermi_a} show how the time derivatives of correlation tensors ${\bf Z}^{(p,\alpha)}$ transform under the action of the super-operators $\mch{F}^{(m,n)}$ and  $\mch{F}^{(m,n)}\mch{P}$ appearing in the fermionic Liouvillian \eref{eq:liouv_fermi}.
The left and the right diagram have a different sign of $m$ and $n$ on the right hand side of the diagrams. This is a consequence of an additional phase super-operator $\mch{P}$ in ${\bf Z}^{p,1}$, which satisfies the following relations
\begin{eqnarray}
\mch{P}f_{j}^+=-f_{N+j}\mch{P},\quad \mch{P}f_{N+j}^+=-f_{j}\mch{P},\quad \mch{P}^2=\ds{1}.
\end{eqnarray}
Hence, the super-operators $\mch{F}^{(m,n)}$, which lower the order of ${\bf Z}^{(p,0)}$, increase the order of ${\bf Z}^{(p,1)}$. Observing this we find three {\it fermionic closeness conditions} for the time evolution of the correlation tensors ${\bf Z}^{(p,\alpha)}$:
\begin{itemize}
\item $m+n$ : even and $n-m\leq 0$ for all non-vanishing $\mch{F}^{(m,n)}$ in \eref{eq:liouv_fermi} ; $\alpha=0$,
\item $m+n$ : even and $m-n\leq 0$ for all non-vanishing $\mch{F}^{(m,n)}$ in \eref{eq:liouv_fermi} ; $\alpha=1$,
\item $\Big(m-n=K$, where $K$ is an odd integer for all non-vanishing $\mch{F}^{(m,n)}\mch{P}$ in \eref{eq:liouv_fermi}  with odd $m+n$\Big) or \Big($m-n=0$ for all non-vanishing  $\mch{F}^{(m,n)}$ in \eref{eq:liouv_fermi} with even $m+n$\Big); $\alpha=0,1$.
\end{itemize}
With $\alpha$ we mark which correlations ${\bf Z}^{(p,\alpha)}$ exhibit a closed hierarchy.  The first condition has been exploited to find steady states of dissipative one-dimensional noninteracting spin system \cite{Pro08,ZP10,Pro10,PZ10}, the XX model with dephasing \cite{Zni10}, and the quantum analog of simple symmetric exclusion process \cite{TWV12}. 

Again we first independently consider the closeness conditions for unitary and non-unitary generators. In case of unitary evolution we find that the closeness condition cannot be satisfied with non-quadratic (or linear) Hamiltonians (see the \Aref{closeness_hamilton}).  In case of a dissipative evolution defined with the dissipator \eref{eq:general_D} and the fermionic coupling operator basis 
\begin{eqnarray}
\label{coupling_fermion}
L_{\ul{j},\ul{k}}=c^\dag_{j_1}c^\dag_{j_2}\ldots c^\dag_{j_\mu} c_{k_1}c_{k_2}\ldots c_{k_\nu},
\end{eqnarray}
the closeness condition implies additional symmetries of the matrix $\bf{G}$. The procedure to attain these symmetries is outlined in the \Aref{closeness_diss}. We explicitly provide only the "lowest order" constraints, which are satisfied by the rate matrix with the symmetries
\begin{eqnarray}
\label{symm_quad_fermion}
G^{(\mu,\nu;\mu',\nu')}_{\ul{j},\ul{k};\ul{j'},\ul{k'}} +\Phi(\mu,\nu,\mu'\nu')\bar{G}^{(\nu,\mu;\nu',\mu')}_{\ul{k},\ul{j};\ul{k'},\ul{j'}} =0,
\end{eqnarray}
where we define the phase factor $\Phi(\mu,\nu,\mu'\nu')=(-1)^{\lfloor \frac{\mu+\nu}{2}\rfloor+\lfloor \frac{\mu'+\nu'}{2}\rfloor+1 + \nu \nu' + \mu' \nu + \mu' \nu' + \mu \mu' + \mu \nu + \mu \nu'}$. In the simplest nontrivial example with the coupling operator basis $L_{j,k}\in\{c_j^\dag c_k,c_j^\dag c_k^\dag, c_j c_k\}$ the symmetry \eref{symm_quad_fermion}  is also sufficient for the first fermionic closeness condition to be satisfied. Conditions \eref{symm_quad_fermion} extend the result of \cite{Eis11}, where it was shown that Hermitian Lindblad operators of the form $L=\sum_{j,k}A_{jk}c_j^\dag c_k$ are a sufficient for closed hierarchy of correlations. In fact by using similar arguments as in the bosonic case for the dissipator \eref{eq:noise} we can prove that in this simple case Hermitian ${\bf A}$ is also a necessary condition for the closeness of the hierarchy. However, it is not clear if \eref{symm_quad_fermion} is the only solution to the conditions for the general quadratic noise presented in the \Aref{closeness_diss}.
\subsection{Exact solution of a dissipative Fermi-Hubbard model}
\label{sec:Hubbard} We argued that interacting fermionic models do not satisfy the fermionic closeness conditions whereas the "interacting" fermionic dissipation does. Here we shall show that adding controlled dissipation can lead to closed hierarchy of equation for the Fermi-Hubbard model with the Hamiltonian
\begin{eqnarray}
\label{Hubbard_H}
H_{\rm Hub}=\sum_{j,k=1}^N t_{jk}c_{j,\sigma}^\dag c_{k\sigma}+ \sum_{j=1}^NU_j c^\dag_{j,\uparrow}c^\dag_{j,\downarrow}c_{j,\uparrow}c_{j,\downarrow}.
\end{eqnarray}
The Fermi-Hubbard model introduced in \cite{Gut63,Hub63,Kan63} is the simplest model describing basic concepts of condensed matter physics, as superconductivity and magnetic end electronic properties of materials and can moreover be realized with cold fermi gases in optical lattices \cite{Ess10}. Although the one-dimensional Fermi-Hubbard model is Bethe ansatz solvable \cite{Ess05}  we shall abstain from using the quantum integrability insight into the structure of the model, but will instead employ a controlled dissipation to simplify the treatment of the model in arbitrary dimension, i.e. for arbitrary hopping $t_{j,k}$ and onsite interaction $U_j$ ($j,k=1,2\ldots N$). 

\begin{figure} [!!h]
\subfiguretopcaptrue
\centering{\subfigure[Correlations in a cubic lattice]{\includegraphics[width=0.5\textwidth]{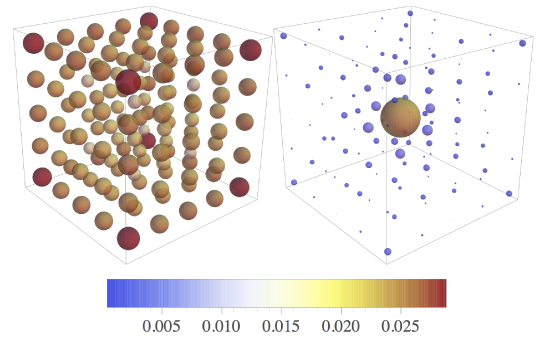}\label{fig:Correlations_a}}
\subfigure[Two-point correlation matrix]{\includegraphics[width=0.38\textwidth]{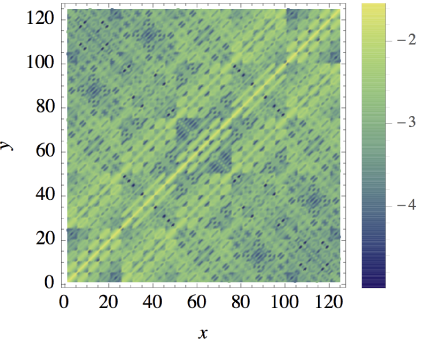} \label{fig:Correlations_c}}}
\caption{a) Occupation number in the steady-state (left) and  absolute value of the correlations (right) $\ave{c_{3,3,3,\uparrow}^\dag c_{i,j,k,\uparrow}}$. The site (3,3,3) is in the middle of the box. b) The two-point correlation matrix $|\ave{c_{i,j,k,\uparrow}^\dag c_{i',j',k',\uparrow}}|$, where $x=i+(j-1)n+(k-1)n^2$, $y=i'+(j'-1)n+(k'-1)n^2$, and $n=5$ is number of sites along one edge. The colour scale in b) is logarithmic. In figures a) the size of the occupations/correlations is proportional to the diameter of the sphere.}
\label{fig:Correlations}
\end{figure}

Since the only term in \eref{Hubbard_H} violating the fermionic closeness conditions is the interacting term we can without loss of generality focus on the Hamiltonian of the form
\begin{eqnarray}
\label{Hub_int}
H_{\rm int}=U c^\dag_{1}c^\dag_{2}c_{1}c_{2}.
\end{eqnarray}
For brevity we omit the sub-index $j$ and define $c_{1,2}^{(\dag)}:=c_{\uparrow,\downarrow}^{(\dag)}$. First, we write the adjoint map of \eref{Hub_int} with aid of the creation and annihilation super-operators
\begin{eqnarray}
\label{adh_Hubb}
\hat{\rm ad}_{H_{\rm int}}=U(&\hat{f}_1^+\hat{f}_4\hat{f}_1\hat{f}_2-\hat{f}_2^+\hat{f}_3\hat{f}_1\hat{f}_2\hat{f}^+_3\hat{f}_3\hat{f}_4\hat{f}_2-\hat{f}^+_4\hat{f}_3\hat{f}_4\hat{f}_1) +\mbox{rest},
\end{eqnarray}
where the rest contains the hierarchy preserving terms. Next, we wish to add dissipation of the form \eref{eq:general_D} with the coupling operator basis $L_{\ul{j},\ul{k}}\in\{c_{1},c_2, c^\dag_{2}c_2 c_1,c^\dag_{1}c_1 c_2, c^\dag_{1}c_2^\dag c_2,c^\dag_{2}c_1^\dag c_1,c_1^\dag,c_2^\dag\}$  which cancels the unwanted terms in \eref{adh_Hubb}. In the \Aref{Sol_Fermi_Hub} we provide a positive Hermitian rate matrix for which the dissipator \eref{eq:general_D} exactly cancels the hierarchy violating terms  such that the complete Liouvillian satisfies the fermionic closeness condition. For this reason it is possible to calculate the dynamics of low order correlations as well as their steady state values for a general Fermi-Hubbard model \eref{Hubbard_H} with a fine-tuned dissipation (see the \Aref{Sol_Fermi_Hub}). In particular we find in one-dimension with nearest neighbour hopping that all steady state two-point correlations vanish aside from the occupation numbers $\ave{c_{j,\sigma}^\dag c_{j,\sigma}}=\frac{G_{55}}{G_{11}-G_{33}+G_{55}}$. In two and three dimensions we were unable to find an analytic expression for the covariance matrix. However, we numerically find exponential decay of correlations with the distance from the diagonal in the two-dimensional case (see the \Aref{Sol_Fermi_Hub}). In three dimensions we observe a revival of correlations at the edges of the cubic lattice, as can be seen in \Fref{fig:Correlations_a} where we show the correlations with respect to the site in the middle of the lattice. We also show in \Fref{fig:Correlations_a} the spatial distribution of the occupations and in \Fref{fig:Correlations_c} the non-vanishing part of the full two-point correlation matrix, which indicates exponential decay of correlations with the distance between sites.  Interestingly, in the described cases the steady state expectation values of the two-point correlations do not depend on the interaction strength. However, their evolution before reaching the steady state and the higher order correlations still do.
A different dissipation can induce interaction-dependent two-point steady-state correlations, and can perhaps reveal some interesting properties of the Fermi-Hubbard model in one or more dimensions.

\section{Mixed Liouvillians}
In this section we determine the mixed closeness conditions. We show that in contrast to the unitary evolution the purely dissipative evolution can exhibit a closed hierarchy. Finally we show an example of dissipation-induced hierarchy of equations for the mixed correlations by providing exact analytic results for time evolution of second order correlations in the dissipative Rabi model.
\subsection{Mixed closeness conditions}
We anew employ the same strategy as for bosons and fermions and rewrite the complete mixed Liouvillian  with $\mch{B}^{(m',n')}$ and $\mch{F}^{(m,n)}$ as defined in \eref{eq:liouv} and \eref{eq:liouv_fermi}, respectively,
\begin{eqnarray}
\mch{L}&=\sum_{m,n,m'n'}\mch{B}^{(m',n')}\mch{F}^{(m,n)}\mch{P}^{m+n}.
\label{mixed_L}
\end{eqnarray}
The fermionic and the bosonic Liouvillian commute, $[\mch{B},\mch{F}]=0$. We define the mixed $(p+q)$-point correlation tensor as
\begin{eqnarray}
\label{corr_mixed}
Z^{(p,q,\alpha)}_{j_1,\ldots,j_p,k_1,\ldots,k_q}(t)&=\brao{1}\hat{b}_{j_1}\ldots\hat{b}_{j_p}\hat{f}_{k_1}\ldots\hat{f}_{k_q}\mch{P}^\alpha\ket{\rho(t)},
\end{eqnarray}
with $\alpha=0,1$ determining the presence of the parity supero-perator in the correlation matrix.
Resembling the fermionic case the two different correlators are necessary due to the presence of the fermionic parity super-operator in the mixed  Liouvillian. The time evolution of the correlators \eref{corr_mixed} is governed by

\begin{eqnarray}
\ddt Z^{(p,q,\alpha)}_{j_1,\ldots j_p,k_1,\ldots k_q}(t)&=\brao{1}\hat{b}_{j_1}\ldots\hat{b}_{j_p}\hat{f}_{k_1}\ldots\hat{f}_{k_q}\mch{P}^\alpha\mch{L}\ket{\rho(t)}.
\label{eq:mixed_Z}
\end{eqnarray}
The action of the mixed Liouvillian \eref{mixed_L} on the correlation tensors can be explained by the following diagrams
\begin{eqnarray}
\label{mixed_diagrams_a}
\xymatrix{
        {\bf Z}^{(p,q,0)} \ar[drrr]|{\mch{B}^{(m',n')}\mch{F}^{(m,n)}\mch{P}} \ar[rrr]^{\!\!\!\!\!\!\!\!\!\!\!\!\!\!\mch{B}^{(m',n')}\mch{F}^{(m,n)}} & &&{\bf Z}^{(p-m'+n',q-m+n,0)} \\
                             &  & & {\bf Z}^{(p-m'+n',q-m+n,1)} } \\ 
            \xymatrix{
        {\bf Z}^{(p,q,1)} \ar[drrr]|{\mch{B}^{(m',n')}\mch{F}^{(m,n)}\mch{P}} \ar[rrr]^{\!\!\!\!\!\!\!\!\!\!\!\!\!\!\mch{B}^{(m',n')}\mch{F}^{(m,n)}} & &&{\bf Z}^{(p-m'+n',q+m-n,1)} \\
                             &  & & {\bf Z}^{(p-m'+n',q+m-n,0)} }\label{mixed_diagrams_b}
\end{eqnarray}
Again due to an additional phase super-operator in the diagram \eref{mixed_diagrams_b}, we have the opposite sign of $m$ and $n$ in comparison to the diagram \eref{mixed_diagrams_a}. On the other hand, the bosonic part (i.e. $m'$ and $n'$) does not change the sign. Taking into account all transitions depicted in the diagrams \eref{mixed_diagrams_a} and \eref{mixed_diagrams_b} we deduce that the equations for the correlators ${\bf Z}^{(p,q,\alpha)}$ are closed under following conditions ({\it mixed closeness conditions}): 
\begin{itemize}
\item $m+n$ : even and $n-m\leq m'-n'$ for all non-vanishing $\mch{F}^{(m,n)}\mch{B}^{(m',n')}$ in \eref{mixed_L} ; $\alpha=0$,
\item $m+n$ : even and $m-n\leq m'-n'$ for all non-vanishing $\mch{F}^{(m,n)}\mch{B}^{(m',n')}$ in \eref{mixed_L} ; $\alpha=1$,
\item $\Big(|m_1-m_2-n_1+n_2|\leq m_1'+m_2'-n_1'-n_2'$ for all non-vanishing paris $\mch{F}^{(m_1,n_1)}\mch{B}^{(m_1',n_1')}\mch{P}$ and $\mch{F}^{(m_2,n_2)}\mch{B}^{(m_2',n_2')}\mch{P}$ in \eref{mixed_L}  with odd $m_1+n_1$ and $m_2+n_2$\Big) or \Big($|m-n|\leq m'-n'$ for  all non-vanishing $\mch{F}^{(m,n)}\mch{B}^{(m',n')}$ in \eref{mixed_L} with even $m+n$\Big); $\alpha=0,1$.
\end{itemize}
If the only non-vanishing  $\mch{F}^{(m,n)}$ in \eref{mixed_L} has $m+n=0$ above conditions reduce to simple bosonic closeness conditions and similarly if  the only non-vanishing $\mch{B}^{(m',n')}$ in \eref{mixed_L}  has $m'+n'=0$ they simplify to fermionic closeness conditions.  Once more we separately consider unitary and dissipative dynamics and observe that the mixed closeness conditions are {\it not} satisfied for any fermion-boson Hamiltonian (see the \Aref{closeness_hamilton}). On the contrary, we can find a fermion-boson dissipator satisfying one of the mixed closeness conditions. The simplest example considered in the \Aref{closeness_diss} is a dissipative evolution determined by \eref{eq:general_D} , with the coupling operator basis $L\in\{c,\,c^\dag,\,a,\,a^\dag \}$ and the rate matrix
\begin{eqnarray}
{\bf G}=\left(
\begin{array}{cccc}
G_{11}&0&G_{13}&0\\
0&G_{11}&0&\bar{G}_{13}\\
\bar{G}_{13}&0&G_{44}+2\delta&0\\
0&G_{13}&0&G_{44}
\end{array}
\right).
\end{eqnarray}
The positivity of ${\bf G}$ implies $|G_{13}|<G_{11}G_{44}$ and $G_{11},G_{44}>0$, whereas the stability of the system implies $\delta>0$.

\subsection{Exact solution of a dissipative Rabi model}
In previous sections we showed that in contrast to unitary evolution purely dissipative evolution satisfies the closeness condition also for non-quasi-free higher-order Liouvillians and can moreover induce a closed hierarchy in case of interacting fermionic and bosonic Hamiltonians. Now we extend this result and show that controlled coupling to a Markovian environment can induce a closed hierarchy of equations also in case of  interacting mixed Hamiltonians. For this purpose, we study the Rabi model \cite{Rab35}, which describes the interaction between a two level system (spineless fermion) and a quantized harmonic oscillator (boson), and is given by the Hamiltonian
\begin{eqnarray}
\label{Rabi_H}
H_{\rm Rabi}=\omega a^\dag a+\Delta c^\dag c + g (c+c^\dag)(a+a^\dag).
\end{eqnarray}
Despite its simplicity it displays a rich behavior and has a wide range of applicability in quantum optics \cite{Ved05}, quantum information \cite{RBH01}, and condensed matter \cite{Hol59}.  Moreover, the time evolution of interesting observables is still restricted to numerical \cite{WDD+08} and analytical \cite{Iri07} approximations. Here we show how the model can be simplified by adding Markovian dissipation which leads to a closed hierarchy of equations for the correlation tensors. We provide exact closed-form expressions for the time evolution of low order correlations. 

Let us first start by writing the action of the adjoint map of the Hamiltonian in terms of the fermionic and bosonic creation and annihilation super-operators
\begin{eqnarray}
\label{adHRabi}
\hat{\rm ad}_{H_{\rm Rabi}}=&\omega(\hat{b}_1^+\hat{b}_1-\hat{b}_2^+\hat{b}_2)+\Delta(\hat{f}_1^+\hat{f}_1-\hat{f}_2^+\hat{f}_2)\\ \nonumber
&+\frac{g}{\sqrt{2}}\Big(\hat{b}^+_1(\hat{f}_1-\hat{f}_2+\hat{f}_2^++\hat{f}_1^+)\\ \nonumber
&-\hat{b}^+_2(\hat{f}_1-\hat{f}_2+\hat{f}_1^+-\hat{f}_2^+)+2(\hat{b}_1+\hat{b}_2)(\hat{f}^+_2-\hat{f}_1^+)\Big)\mch{P}.
\end{eqnarray}
We observe that the closeness conditions are not satisfied due to the terms of the form $\hat{f}^+_j\hat{b}_k\mch{P}$, where $j,k=1,2$. The idea is to add noise (dissipation) which exactly cancels these terms. Again we employ the dissipator of the form \eref{eq:general_D} with the coupling operator basis $ L\in \{c,\,c^\dag,\, a,\,a^\dag\}$ and rate matrix
\begin{eqnarray}
\label{solRabi}
{\bf G}=\left(\begin{array}{cccc}
G_{11} & G_{1,2} & 0 & -2\,\ii \, g \\
G_{21} & G_{22} & 0 & -2\,\ii \, g \\
0 & 0 & G_{33} & G_{3,4} \\
2\,\ii \,g & 2\,\ii\,g & G_{43} & G_{44} \\
\end{array}\right).
\end{eqnarray}
This dissipator exactly cancels the unwanted terms in \eref{adHRabi}. Moreover, the so far unspecified rates $G_{i,j}$ in \eref{solRabi} can be chosen such that ${\bf G}$ is a positive Hermitian matrix and consequently defines a valid Lindblad dissipator. For simplicity we choose $G_{11}=G_{22}=G_{44}=\Gamma>2 \sqrt{2}|g|$, $G_{33}=\Gamma+2\delta>0$, and $G_{2,1}=G_{1,2}=G_{3,4}=G_{4,3}=0$. In this case the eigenvalues of ${\bf G}$  are positive ($\Gamma, \Gamma+2\delta, \Gamma- 2\sqrt{2}\,g, \Gamma + 2\sqrt{2}\,g$). The full Liouvillian $\mch{L}_{\rm Rabi}=-\ii\,\hat{\rm ad}_{H_{\rm Rabi}}+\mch{D}_{\rm Rabi}$ expressed in terms of bosonic \eref{eq:mb} and fermionic \eref{eq:mf} creation and annihilation maps is
\begin{eqnarray}
\label{LRabi}
\!\!\!\!\!\!\!\!\!\!\!\!\!\!\!\!\!\!\!\!\!\!\!\!\!\!\!\!\!\!\!\!\!\!\!\!\mch{L}_{\rm Rabi}=&-\ii\Bigg( \omega(\hat{b}_1^+\hat{b}_1-\hat{b}_2^+\hat{b}_2)+\Delta(\hat{f}_1^+\hat{f}_1-\hat{f}_2^+\hat{f}_2)\\ \nonumber
\!\!\!\!\!\!\!\!\!\!\!\!\!\!\!\!\!\!\!\!\!\!\!\!\!\!\!\!\!\!\!\!\!\!\!\!& ~~~~~~+\sqrt{2}  \,g\, \Big(\hat{b}_1^+ \hat{f}_1  - \hat{b}_1^+ \hat{f}_2 -  \hat{b}_2^+ \hat{f}_1 + \hat{b}_2^+ \hat{f}_2  + \hat{b}_1^+\hat{f}_1^+ - \hat{b}_1^+ \hat{f}_2^++  \hat{b}_2^+ \hat{f}^+_1 - \hat{b}_2^+ \hat{f}_2^+\Big)\mch{P}\Bigg)\\ \nonumber
\!\!\!\!\!\!\!\!\!\!\!\!\!\!\!\!\!\!\!\!\!\!\!\!\!\!\!\!\!\!\!\!\!\!\!\!&+\Gamma(\hat{b}^+_1\hat{b}^+_2-\hat{f}_1^+\hat{f}_1-\hat{f}_2^+\hat{f}_2)-\delta(\hat{b}_1^+\hat{b}_1+\hat{b}_2^+\hat{b}_2).
\end{eqnarray}
Since the Liouvillian \eref{LRabi} does not contain any bosonic annihilation maps we have a closed hierarchy of equations for the correlation tensors, which are easily solvable for low order correlations. In particular we find the following expressions for correlations of the first and second order
\begin{eqnarray}
\ave{c}&=&\ave{a}=\ave{c^\dag c-cc^\dag}=0 \\ \nonumber
\ave{ac}&=&\frac{2\, \ii \,g \left(-1+e^{-t (\Gamma +\delta +\ii (\Delta +\omega
   ))}\right)}{\Gamma +\delta +\ii (\Delta +\omega )},\\ \nonumber
 \ave{a^\dag c}&=&\frac{- 2\,\ii\, g \left(-1+e^{-t (\Gamma +\delta +\ii (\Delta -\omega
   ))}\right)}{\Gamma +\delta +\ii (\Delta-\ii\omega) },\\ \nonumber
\end{eqnarray}
Since the remaining expressions for the time evolution of correlations are too lengthy we provide only their steady state values.
\begin{eqnarray}
\ave{a^\dag a}&=&\frac{\Gamma +\frac{4 g^2 (\Gamma +\delta )}{(\Gamma +\delta )^2+(\Delta
   -\omega )^2}+\frac{4 g^2 (\Gamma +\delta )}{(\Gamma +\delta
   )^2+(\Delta +\omega )^2}}{2 \delta },\\ \nonumber
\ave{a^2}&=&-\frac{4 g^2 (\Gamma +\delta +\ii \omega )}{(\delta +\ii \omega ) (\Gamma
   +\delta -\ii (\Delta -\omega )) (\Gamma +\delta +\ii (\Delta +\omega ))}.
\end{eqnarray}
The system is initially (at time $t=0$) in bosonic vacuum and fermionic totally mixed state. The system is stable if $\delta>0$, i.e. when more bosonic modes are incoherently dissipated than pumped into the system. We may include also coherent pumping or more fermions or bosons and treat the model in a similar manner.

\section{Conclusions}
We determined simple criteria for the closeness of the hierarchy of equations in Markovian open quantum systems. In particular we discussed general quadratic fermionic and bosonic noise. We showed that dissipative systems satisfy the closeness property under more general conditions than closed systems. In case of unitary evolution the closeness of the hierarchy is possible only in quadratic systems, in other words for an evolution which preserves Gaussian states. In the dissipative case, however, the closeness conditions can be satisfied by more general systems, of which steady states are not necessary Gaussian states. Nevertheless, their low order correlations (as well as their dynamics before reaching the steady state) can be efficiently computed. Moreover, we considered three examples of dissipation-induced closeness of the hierarchy, where the unitary evolution, determined by interacting Hamiltonians, by itself does not satisfy the closeness condition, whereas the joint non-unitary evolution does. This enabled us to find exact expressions for the time evolution of the first and second order correlations of paradigmatic interacting quantum models, which are otherwise accessible only through numerical approximations. Although the solutions may seem rather artificial due to an additional fine-tuned noise they can perhaps provide new insight into the structure of interacting quantum models in one and more dimensions and can be extended by means of perturbation theory in the interaction strength. 

Albeit the closeness conditions enable simple calculations of low order correlations a general procedure how to evaluate higher order correlations, as the Wick's theorem for Gaussian systems, remains an interesting open problem. Since the closeness condition also enables the calculation of some eigenvalues of the Liouvillian it would be interesting to see if this eigenvalues include also the spectral gap, which is the eigenvalue with the largest real part and generically governs the convergence of the evolution to the steady state.

\section*{Acknowledgments}
The author thanks S. Ajisaka, F. Barra, E. Ilievski, and T. Prosen, for reading the manuscript and useful comments. The work was supported by  FONDECYT project 3130495.

\section*{References}
\bibliographystyle{ieeetr}
\bibliography{Bibliography}

\appendix

\section{Closed hierarchy for unitary dynamics}
\label{closeness_hamilton}
In this Appendix we study the property of the closeness of correlations for fermionic, bosonic, and mixed Hamiltonians. 
\subsection{Bosonic Hamiltonians}
A general bosonic Hamiltonian can be written in the form
\begin{eqnarray}
\label{boson_hamiltonian}
H=\sum_{\mu,\nu}\sum_{\underline{j}\underline{k}}H_{\underline{j},\underline{k}}^{(\mu,\nu)}a^\dag_{j_1}a^\dag_{j_2}\ldots a^\dag_{j_m}a_{k_1}a_{k_2}\ldots d_{k_n},
\end{eqnarray}
where the indices $\mu$ and $\nu$ determine the lengths of the vectors $\ul{j}$ and $\ul{k}$, respectively. It is clear that due to the trace preservation and the property $\bra{\mathds{1}}\hat{b}^+=0$  linear and quadratic Hamiltonians automatically satisfy the closeness condition. In order to show that higher order Hamiltonians do not respect this rule we look at terms of the adjoint map $\hat{\rm ad}_H\,\rho =[H,\rho]$ that contain only one bosonic creation map $\hat{b}_{j_p}^+$  or $\hat{b}_{N+j_p}^+$  
\begin{eqnarray}
\label{adh_map_boson}
\hat{\rm ad}_H=&\sum_{\mu,\nu}\sum_{p=1}^{\mu}\sum_{\underline{j},\underline{k}}H^{(\mu,\nu)}_{\underline{j},{\underline{k}}}\hat{b}_{j_p}^+\prod_{i=1,i\neq p}^\mu\hat{b}_{N+j_i}\prod_{i=1}^{\nu}\hat{b}_{k_i}\\ \nonumber
&-\sum_{\mu,\nu}\sum_{p=1}^{\nu}\sum_{\underline{j},\underline{k}}H^{(\mu,\nu)}_{\underline{j},{\underline{k}}}\hat{b}_{N+k_p}^+\prod_{i=1,i\neq p}^{\nu}\hat{b}_{k_i}\prod_{i=1}^{\mu}\hat{b}_{N+j_i}+{\rm rest}.
\end{eqnarray}
The rest contains all remaining terms, which have two or more bosonic creation maps or in which the number of creation and annihilation maps is not equal to $\mu+\nu$. One can immediately see that due to the symmetry $H_{P\underline{j},P'\underline{k}}=H_{\underline{j},\underline{k}}$, where $P$ and $P'$ represent arbitrary permutations of elements of vectors $\underline{j}$ and $\underline{k}$, respectively, the explicit sums in equation \eref{adh_map_boson} cannot vanish. Therefore, the map $\hat{\rm ad}_H$ always contains terms with only one creation super-operator and $\nu+\mu-1$ annihilation super-operators, which leads to the following closeness condition $M_{\rm max}=\max(\nu+\mu-2)\leq0$, where the maximum is taken over all pairs $\mu,\nu$ in \eref{boson_hamiltonian} and \eref{adh_map_boson} for which ${\bf H}^{(\mu,\nu)}$ does not vanish. In other words, the closeness condition for bosonic Hamiltonians is satisfied only for quadratic and linear bosonic Hamiltonians.
\subsection{Fermionic Hamiltonians}
In this subsection we discuss the closeness conditions for general fermionic Hamiltonians
\begin{eqnarray}
\label{fermi_H}
H=\sum_{\mu,\nu}\sum_{\underline{j}\underline{k}}H_{\underline{j},\underline{k}}^{(\mu,\nu)}c^\dag_{j_1}c^\dag_{j_2}\ldots c^\dag_{j_\mu}c_{k_1}c_{k_2}\ldots c_{k_\nu}.
\end{eqnarray}
Explicitly writing only terms with exactly one fermionic creation map $\hat{f}_{j_p}^+$  or $\hat{f}_{N+j_p}^+$ the adjoint map of the fermionic Hamiltonian \eref{fermi_H} reads
\begin{eqnarray}
\label{adh_map_fermion}
\hat{\rm ad}_H=&\sum_{\mu,\nu}\sum_{p=1}^{\mu}\frac{(-1)^{\lfloor (\nu+\mu)/2\rfloor+\mu+p+1}}{2^{(\nu+\mu)/2}}\sum_{\underline{j},\underline{k}}H^{(\mu,\nu)}_{\underline{j},{\underline{k}}}\hat{f}_{j_p}^+\prod_{i=1;i\neq p}^\mu\hat{f}_{N+j_i}\prod_{i=1}^{\nu}\hat{f}_{k_i}\mch{P}^{\mu+\nu}\\ \nonumber
&-\sum_{\mu,\nu}\sum_{p=1}^{\mu}\frac{(-1)^{\mu-p }}{2^{(\nu+\mu)/2}}\sum_{\underline{j},\underline{k}}H^{(\mu,\nu)}_{\underline{j},{\underline{k}}}\prod_{i=\nu}^{1}\hat{f}_{k_i}\prod_{i=\mu;i\neq p}^1\hat{f}_{N+j_i} \, \hat{f}_{j_p}^+\mch{P}^{\mu+\nu}\\ \nonumber
&+\sum_{\mu,\nu}\sum_{p=1}^{\nu}\frac{(-1)^{\lfloor (\mu+\nu)/2\rfloor+\mu+\nu-p }}{2^{(\nu+\mu)/2}}\sum_{\underline{j},\underline{k}}H^{(\mu,\nu)}_{\underline{j},{\underline{k}}}\prod_{i=1}^{\mu}\hat{f}_{N+j_i}\prod_{i=1;i\neq p}^\nu\hat{f}_{k_i} \, \hat{f}_{N+k_p}^+\mch{P}^{\mu+\nu}\\ \nonumber
&-\sum_{\mu,\nu}\sum_{p=1}^{\nu}\frac{(-1)^{\nu+\mu-p+1}}{2^{(\nu+\mu)/2}}\sum_{\underline{j},\underline{k}}H^{(\mu,\nu)}_{\underline{j},{\underline{k}}}\hat{f}_{N+k_p}^+\prod_{i=\nu;i\neq p}^1\hat{f}_{k_i}\prod_{i=\mu}^{1}\hat{f}_{N+j_i}\mch{P}^{\mu+\nu}\\ \nonumber 
&+{\rm rest}.
\end{eqnarray}
The rest contains all terms which either have two or more creation maps or consist of less than $\mu+\nu$ creation and annihilation maps. By changing the order of operators in the second an fourth sum in \eref{adh_map_fermion} we obtain
\begin{eqnarray}
\label{adh_fermi1}
\hat{\rm ad}_H=&\sum_{\mu,\nu}\sum_{p=1}^{\mu}\frac{(-1)^{\lfloor (\nu+\mu)/2\rfloor+\mu+p+1}}{2^{(\nu+\mu)/2-1}}\sum_{\underline{j},\underline{k}}H^{(\mu,\nu)}_{\underline{j},{\underline{k}}}\hat{f}_{j_p}^+\prod_{i=1;i\neq p}^\mu\hat{f}_{N+j_i}\prod_{i=1}^{\nu}\hat{f}_{k_i}\mch{P}^{\mu+\nu}\\ \nonumber
&+\sum_{\mu,\nu}\sum_{p=1}^{\nu}\frac{(-1)^{\lfloor (\mu+\nu)/2\rfloor+\mu+\nu-p}}{2^{(\nu+\mu)/2-1}}\sum_{\underline{j},\underline{k}}H^{(\mu,\nu)}_{\underline{j},{\underline{k}}}\prod_{i=1}^{\mu}\hat{f}_{N+j_i}\prod_{i=1;i\neq p}^\nu\hat{f}_{k_i} \, \hat{f}_{N+k_p}^+\mch{P}^{\mu+\nu}\\ \nonumber&+{\rm rest}.
\end{eqnarray}

Using the symmetry $H_{P\underline{j},P'\underline{k}}=\varepsilon_{P\ul{j}}\varepsilon_{P'\ul{k}}H_{\underline{j},\underline{k}}$, with $P$ and $P'$ being arbitrary permutations of elements of vectors $\underline{j}$ and $\underline{k}$, respectively, and $\varepsilon_{\underline{j}}$ a completely antisymmetric tensor, we observe that terms with one creation map and $\mu+\nu-1$ annihilation maps do not vanish. Hence, for even $\mu+\nu$ the first fermionic closeness condition reduces to $\nu+\mu-2\leq0$. In order to show that the second and the third condition can not be satisfied for $\mu+\nu>2$, we examine other terms with maximal number of creation and annihilation maps. Let us first write out only the terms consisting only of creation maps
\begin{eqnarray}
\label{adh_fermi2}
\!\!\!\!\!\!\!\!\!\!\!\!\!\!\!\!\!\!\!\!\!\!\!\!\!\!\!\!\!\!\!\!\!\!\!\!\hat{\rm ad}_H=&\sum_{\mu,\nu}\sum_{p=1}^{\mu}\frac{(-1)^{\lfloor (\nu+\mu)/2\rfloor+\mu+p+1}}{2^{(\nu+\mu)/2-1}}\sum_{\underline{j},\underline{k}}H^{(\mu,\nu)}_{\underline{j},{\underline{k}}}(1-(-1)^{\mu+\nu})\prod_{i=1}^\mu\hat{f}_{N+j_i}\prod_{i=1}^{\nu}\hat{f}_{k_i}\mch{P}^{\mu+\nu}\\ \nonumber
\!\!\!\!\!\!\!\!\!\!\!\!\!\!\!\!\!\!\!\!\!\!\!\!\!\!\!\!\!\!\!\!\!\!\!\!&+{\rm rest},
\end{eqnarray}
where the rest contains all non-vanishing terms with at least one annihilation map. For odd $\mu+\nu$ explicit terms in \eref{adh_fermi2} do not vanish. Due to \eref{adh_fermi1} and \eref{adh_fermi2} the fermionic adjoint map \eref{eq:liouv_fermi} with odd $\mu+\nu$ always contains terms with $n-m<0$ and terms with $m>1$ and $n=0$ and can {\it not} satisfy the third fermionic closeness condition unless $\mu+\nu=1$. Since for even $\mu+\nu$ the explicit terms in \eref{adh_fermi2} vanish we have to consider terms of the adjoint map containing only one annihilation map and $\mu+\nu-1$ creation maps. After a tedious calculation we find the following simplified expression
\begin{eqnarray}
\nonumber
\!\!\!\!\!\!\!\!\!\!\!\!\!\!\!\!\!\!\!\!\!\!\!\!\!\!\!\!\!\!\!\!\!\!\!\!\hat{\rm ad}_H&=\sum_{\mu,\nu}\sum_{p=1}^{\mu}\frac{(-1)^{\lfloor (\nu+\mu)/2\rfloor+\mu+p+1}}{2^{(\nu+\mu)/2}}(1+(-1)^{\mu+\nu})\sum_{\underline{j},\underline{k}}H^{(\mu,\nu)}_{\underline{j},{\underline{k}}}\hat{f}_{N+j_p}\prod_{i=1;i\neq p}^\mu\hat{f}_{j_i}^+\prod_{i=1}^{\nu}\hat{f}_{N+k_i}^+\mch{P}^{\mu+\nu}\\ \nonumber
\!\!\!\!\!\!\!\!\!\!\!\!\!\!\!\!\!\!\!\!\!\!\!\!\!\!\!\!\!\!\!\!\!\!\!\!&+\sum_{\mu,\nu}\sum_{p=1}^{\nu}\frac{(-1)^{\lfloor (\mu+\nu)/2\rfloor+\mu+\nu-p}}{2^{(\nu+\mu)/2}}(1+(-1)^{\mu+\nu})\sum_{\underline{j},\underline{k}}H^{(\mu,\nu)}_{\underline{j},{\underline{k}}}\prod_{i=1}^{\mu}\hat{f}_{j_i}^+\prod_{i=1;i\neq p}^\nu\hat{f}_{N+k_i} \, \hat{f}_{k_p}\mch{P}^{\mu+\nu}\\ \nonumber
\!\!\!\!\!\!\!\!\!\!\!\!\!\!\!\!\!\!\!\!\!\!\!\!\!\!\!\!\!\!\!\!\!\!\!\!&+{\rm rest}\label{adh_fermi3}
\end{eqnarray}
implying that the second fermionic condition can {\it not} be satisfied unless $\mu+\nu=2$. We showed that similarly as in the bosonic case the only two fermionic Hamiltonians with the closed hierarchy of equations for correlations are quadratic and linear Hamiltonians. Since explicit terms in \eref{adh_fermi2} vanish the Hamiltonian with quadratic and linear terms also satisfies the third fermionic closeness condition.
\subsection{Mixed Hamiltonians}
In this subsection we determine the closeness condition for the mixed Hamiltonian of the form
\begin{eqnarray}
\label{mix_H}
H=\sum_{\mu,\nu,\mu',\nu'}\sum_{\ul{j},\ul{k},\ul{t},\ul{v}}H^{(\mu,\nu,\mu',\nu')}_{\ul{j},\ul{k},\ul{t},\ul{v}}c^\dag_{j_1}\ldots c^\dag_{j_\mu}c_{k_1}\ldots c_{k_\nu}a^\dag_{t_1}\ldots a^\dag_{t_\mu}a_{v_1}\ldots a_{v_\nu},
\end{eqnarray}
with $\mu+\nu\geq1$ and $\mu'+\nu'\geq1$. The fermionic and the bosonic maps commute. Accordingly, the adjoint map of $\eref{mix_H}$ contains the same fermionic terms as in \eref{adh_fermi1}, \eref{adh_fermi2} and \eref{adh_fermi3} multiplied by $\mu'+\nu'$ bosonic annihilation maps. Therefore, Hamiltonians containing fermion-boson interaction can {\it not} satisfy the closeness conditions.

\section{Closed hierarchy for Lindblad dissipators}
\label{closeness_diss}
In this Appendix we discuss Lindblad dissipators of the form 
\begin{eqnarray}
\label{GeneralD}
&\mch{D}\rho=\sum_{\mu,\nu,\mu',\nu'}\sum_{\ul{j},\ul{k};\ul{j}',\ul{k}'}G^{(\mu,\nu;\mu',\nu')}_{\ul{j},\ul{k};\ul{j}',\ul{k}'}\left(L_{\ul{j},\ul{k}}\rho L_{\ul{j}',\ul{k}'}^\dag-\frac{1}{2}\{L_{\ul{j}',\ul{k}'}^{\dag} L_{\ul{j},\ul{k}},\rho\}\right),
\end{eqnarray}
where $\bf{G}$ is a positive Hermitian matrix rate-matrix, $\mu,\,\nu,\,\mu'\,\,\nu'$ designate the lengths of multi-indices  $\ul{j},\,\ul{k},\,\ul{j}',\,\ul{k}'$ (e.g., $\ul{j}=(j_1,j_2\ldots,j_\mu)$, and  $L_{\ul{j}\ul{k}}$ designates the coupling operators, which shall be defined in the following sections. Our aim is to find necessary and sufficient conditions for the closed hierarchy of equations for correlations \eref{eq:ref z}, \eref{fermion_Z} and \eref{eq:mixed_Z}. We separately study bosonic, fermionic and mixed case. For quadratic bosonic and fermionic Lindblad operators we determine a set of symmetry conditions for ${\bf G}$ which is equivalent to the closeness condition. In the mixed case we study the simplest nontrivial example and show that in contrast to the unitary case the closeness conditions can be satisfied. 
\subsection{Bosons}
\label{adh_bosons}
 In general the bosonic coupling operators can be written as
\begin{eqnarray}
L_\alpha=a^\dag_{j_1}a^\dag_{j_2}\ldots a^\dag_{j_m} a_{k_1}a_{k_2}\ldots a_{k_n}.
\end{eqnarray}  
We take the same approach as in the unitary case. We  first write the dissipator by using creation and annihilation maps and then consider terms with different number of creation maps. From this expansion we extract independent, necessary requirements for the closeness condition. Let us first focus on terms containing only one creation map and $m+n+m'+n'-1$ annihilation maps
\begin{eqnarray}
\mch{D}=&\frac{1}{2}\sum_{\ul{j},\ul{k},\ul{t},\ul{v}}G^{(\mu,\nu;\mu',\nu')}_{\ul{j},\ul{k};\ul{t},\ul{v}}\left(\sum_{p=1}^\mu \hat{b}_{j_p}^\dag \prod_{i=1}^\nu\hat{b}_{k_i}\prod_{i=1}^{\mu'}\hat{b}_{t_i}\prod_{i=1,i\neq p}^\mu\hat{b}_{N+j_i}\prod_{i=1}^{\nu'}\hat{b}_{N+v_i}\right. \\ \nonumber
&- \sum_{p=1}^{\nu'} \hat{b}_{v_p}^\dag \prod_{i=1}^\nu\hat{b}_{k_i}\prod_{i=1}^{\mu'}\hat{b}_{t_i}\prod_{i=1}^\mu\hat{b}_{N+j_i}\prod_{i=1, i\neq p}^{\nu'}\hat{b}_{N+v_i} \\ \nonumber
&+ \sum_{p=1}^{\mu'} \hat{b}_{N+t_p}^\dag \prod_{i=1}^\nu\hat{b}_{k_i}\prod_{i=1,i\neq p}^{\mu'}\hat{b}_{t_i}\prod_{i=1}^\mu\hat{b}_{N+j_i}\prod_{i=1}^{\nu'}\hat{b}_{N+v_i} \\ \nonumber
&\left.- \sum_{p=1}^\nu \hat{b}_{N+k_p}^\dag \prod_{i=1 ,i\neq p}^\nu\hat{b}_{k_i}\prod_{i=1}^{\mu'}\hat{b}_{t_i}\prod_{i=1}^\mu\hat{b}_{N+j_i}\prod_{i=1}^{\nu'}\hat{b}_{N+v_i} \right)+{\rm rest}.
\end{eqnarray}
Since the tensor $\bf{G}$ is symmetric with respect to permutations of elements of each multi index $\ul{j},\ul{k},\ul{t},\ul{v}$ we obtain the conditions
\begin{eqnarray}
\label{diss_boson1}
\!\!\!\!\!\!\!\!\!\!\!\!\!\!\!\!\!\!\sum_{\ul{j},\ul{k},\ul{t},\ul{v}}\left(G^{(\mu,\nu;\mu',\nu')}_{\ul{j},\ul{k};\ul{t},\ul{v}}-G^{(\nu',\nu;\mu',\mu)}_{\ul{v},\ul{k};\ul{t},\ul{j}}\right)\sum_{p=1}^{\mu} \hat{b}_{j_p}^\dag \prod_{i=1}^{\nu}\hat{b}_{k_i}\prod_{i=1}^{\mu'}\hat{b}_{t_i}\prod_{i=1 , i\neq p}^{\mu}\hat{b}_{N+j_i}\prod_{i=1}^{\nu'}\hat{b}_{N+v_i} =0, \\ \nonumber
\!\!\!\!\!\!\!\!\!\!\!\!\!\!\!\!\!\! \sum_{\ul{j},\ul{k},\ul{t},\ul{v}}\left(G^{(\mu,\nu;\mu',\nu')}_{\ul{j},\ul{k};\ul{t},\ul{v}}-G^{(\mu,\mu';\nu,\nu')}_{\ul{j},\ul{t};\ul{k},\ul{v}}\right)\sum_{p=1}^{\nu} \hat{b}_{N+k_p}^\dag \prod_{i=1 , i\neq p}^{\nu}\hat{b}_{k_i}\prod_{i=1}^{\mu'}\hat{b}_{t_i}\prod_{i=1}^{\mu}\hat{b}_{N+j_i}\prod_{i=1}^{\nu'}\hat{b}_{N+v_i} =0.
\end{eqnarray}
Now we introduce the vectors $\ul{x}=(\ul{j},\ul{v})$ and $\ul{y}=(t_2,t_3\ldots,t_{m'},k_1,k_2,\ldots k_{n})$ and extract from the equation \eref{diss_boson1} the following symmetry conditions for the rate matrix~${\bf G}$
\begin{eqnarray}
\label{diss_boson2}
0&=\sum_{\mu+\nu'=n_x}\sum_{\nu+\mu'=n_y}\sum_{P \ul{x},P'\ul{y}}\mu\left( G^{(\mu,\nu;\mu',\nu')}_{\ul{j},\ul{k};\ul{t},\ul{v}}-G^{(\mu,\mu';\nu,\nu')}_{\ul{j},\ul{t};\ul{k},\ul{v}} \right)\\ \nonumber
&=\sum_{\mu+\nu'=n_x}\sum_{\nu+\mu'=n_y}\sum_{P \ul{x},P'\ul{y}}\mu\left( G^{(\mu,\nu;\mu',\nu')}_{\ul{j},\ul{k};\ul{t},\ul{v}}-G^{(\nu',\mu';\nu,\mu)}_{\ul{v},\ul{t};\ul{k},\ul{j}} \right)\\ \nonumber
&=\sum_{\mu+\nu'=n_x}\sum_{\nu+\mu'=n_y}\sum_{P \ul{x},P'\ul{y}}\mu\left( G^{(\mu,\nu;\mu',\nu')}_{\ul{j},\ul{k};\ul{t},\ul{v}}-\bar{G}^{(\nu,\mu;\nu',\mu')}_{\ul{k},\ul{j};\ul{v},\ul{t}} \right),
\end{eqnarray}
which have to be satisfied for all variations of $\ul{x}$, {\ul y} and $t_1$. The simplest solution of the equations \eref{diss_boson2} is if all terms in the sum vanish separately, namely if $G^{(\mu,\nu;\mu',\nu')}_{\ul{j},\ul{k};\ul{t},\ul{v}}-\bar{G}^{(\nu,\mu;\nu',\mu')}_{\ul{k},\ul{j};\ul{v},\ul{t}} $. The question remains if this is in general the only solution which satisfies also the positivity constraint. Equations \eref{diss_boson2} are necessary condition for the closeness of the hierarchy. However, they are sufficient only int the case where $\mu+\nu+\mu'+\nu'=4$. In general we obtain in a similar manner also "higher-order" necessary conditions for the closeness of the hierarchy and it is not clear if they can always be satisfied. Nevertheless, for a general quadratic noise determined by coupling operators from the set $\{a_ja_k,\,a_j^\dag a_k^\dag,\,a^\dag_j a_k\}$ equations \eref{diss_boson2} are sufficient for the closeness of the hierarchy and simplify to
\begin{eqnarray}
\!\!\!\!\!\!\!\!\!\!\!\!\!\!\!\!\!\!\!\!\!\!\!\!\!\!\!\!\!\!\!\!\!0=&  4(G^{(2,0;2,0)}_{j_1,j_2;t_1,t_2}-\bar{G}^{(0,2;0,2)}_{j_1,j_2;t_1,t_2})+ G^{(1,1;1,1)}_{j_1,j_2;t_1,t_2}+G^{(1,1;1,1)}_{t_2,j_2;t_1,j_1}-\bar{G}^{(1,1;1,1)}_{j_2,j_1;t_2,t_1}-\bar{G}^{(1,1;1,1)}_{j_2,t_2;j_1,t_1}, \\ \nonumber
\!\!\!\!\!\!\!\!\!\!\!\!\!\!\!\!\!\!\!\!\!\!\!\!\!\!\!\!\!\!\!\!\!0=& G^{(1,1;2,0)}_{j_1,k_1;t_1,t_2}+G^{(1,1;2,0)}_{j_1,t_2;t_1,k_1}-\bar{G}^{(1,1;0,2)}_{k_1,j_1;t_1,t_2}+G^{(1,1;2,0)}_{t_2,j_1;k_1,t_1}+ G^{(0,2;1,1)}_{j_1,k_1;t_1,t_2}-\bar{G}^{(2,0;1,1)}_{j_1,k_1;t_2,t_1},\\ \nonumber
\!\!\!\!\!\!\!\!\!\!\!\!\!\!\!\!\!\!\!\!\!\!\!\!\!\!\!\!\!\!\!\!\!0=& G^{(0,2;2,0)}_{k_1,k_2;t_1,t_2}-\bar{G}^{(2,0;0,2)}_{k_1,k_2;t_1,t_2}.
\end{eqnarray}
This equations have to bee satisfied for all values of $j_1,j_2,t_1,t_2,k_1,k_2$ and determine the most general set of quadratic Markovian bosonic noise which satisfies the bosonic closeness condition.

\subsection{Fermions}
In this section we study general fermionic dissipators of the form \eref{GeneralD} with the coupling operators
\begin{eqnarray}
L_\alpha=c^\dag_{j_1}c^\dag_{j_2}\ldots c^\dag_{j_m} c_{k_1}c_{k_2}\ldots c_{k_n}.
\end{eqnarray}  
We shall study only the simplest equations leading to necessary and sufficient symmetries of $\bf{G}$ for the first fermionic closeness condition in case of quadratic noise. As noted in the main text, we need to ensure that all terms with nontrivial phase super-operator vanish. After a tedious calculation and bookkeeping of the sign we obtain the following expression for the fermionic dissipator
\begin{eqnarray}
\label{fdissa1}
\!\!\!\!\!\!\!\!\!\!\!\!\!\!\!\!\!\!\!\!\!\!\!\!\!\!\!\mch{D}=\sum_{\ul{j},\ul{k},\ul{t},\ul{v}}&\frac{(-1)^{\lfloor \frac{\mu+\nu}{2} \rfloor+\mu}}{2^{\frac{\mu+\nu+\mu'+\nu'}{2}}}G^{(\mu,\nu;\mu',\nu')}_{\ul{j},\ul{k};\ul{t},\ul{v}} \times\\ \nonumber
&\left(\sum_{p=1}^\mu(-1)^{p+1+\nu'+(m-1)(\mu'+\nu)}\hat{f}_{j_p}^+ \prod_{i=1}^\nu\hat{f}_{k_i}\prod_{i=1}^{\mu'}\hat{f}_{t_i}\prod_{i=1;i\neq p}^\mu\hat{f}_{\nu+j_i}\prod_{i=1}^{\nu'}\hat{f}_{\nu+v_i} \right. \\ \nonumber
&+\sum_{p=1}^{\nu'}(-1)^{p+\mu+\nu+\mu'+\nu'+\mu(\mu'+\nu)}\hat{f}_{v_p}^+ \prod_{i=1}^\nu\hat{f}_{k_i}\prod_{i=1}^{\mu'}\hat{f}_{t_i}\prod_{i=1}^\mu\hat{f}_{\nu+j_i}\prod_{i=1; i\neq p}^{\nu'}\hat{f}_{\nu+v_i} \\ \nonumber
&+\sum_{p=1}^{n}(-1)^{p+1+\mu+\nu'+\mu(\mu'+\nu-1)}\hat{f}_{\nu+k_p}^+ \prod_{i=1;i\neq p}^\nu\hat{f}_{k_i}\prod_{i=1}^{\mu'}\hat{f}_{t_i}\prod_{i=1}^\mu\hat{f}_{\nu+j_i}\prod_{i=1}^{\nu'}\hat{f}_{\nu+v_i} \\ \nonumber
&+\left .\sum_{p=1}^{\mu'}(-1)^{p+\mu+\nu+\nu'+\mu(m'-1+\nu)}\hat{f}_{\nu+t_p}^+ \prod_{i=1}^\nu\hat{f}_{k_i}\prod_{i=1;i\neq p}^{\mu'}\hat{f}_{t_i}\prod_{i=1}^\mu\hat{f}_{\nu+j_i}\prod_{i=1}^{\nu'}\hat{f}_{\nu+v_i} \right)\\ \nonumber
& ~~\times\mch{P}^{\mu'+\nu'+\mu+\nu}+{\rm rest}.
\end{eqnarray}
In order for the first and third fermionic closeness conditions to be satisfied we have to demand that the explicit sums in \eref{fdissa1} vanish. This gives us following conditions
\begin{eqnarray}
\label{fdissa2}
\!\!\!\!\!\!\!\!\!\!\!\!\!\!\!\!\!\!\sum_{\mu+\nu'=n_x}\sum_{\nu+\mu'=n_y}\sum_{P\ul{x},P\ul{y}}\mu(-1)^{1+\mu+(\mu-1)(\mu'+\nu)+\nu'}\epsilon_{\ul{x}}\epsilon_{\ul{y}}\Bigg((-1)^{\lfloor \frac{\mu+\nu}{2}\rfloor} G^{(\mu,\nu;\mu',\nu')}_{\ul{j},\ul{k};\ul{t},\ul{v}}\\ \nonumber
~~~~~~~~~~~~~~~ +(-1)^{\lfloor \frac{\mu'+\nu'}{2}\rfloor+1 + \nu \nu' + \mu' \nu + \mu' \nu' + \mu \mu' + \mu \nu + \mu \nu'}\bar{G}^{(\nu,\mu;\nu',\mu')}_{\ul{k},\ul{j};\ul{v},\ul{t}} \Bigg)=0,
\end{eqnarray}
where $\ul{y}=(j_2,j_3\ldots,j_{m},v_1,v_2,\ldots v_{n'})$,  $\ul{y}=(\ul{k},\ul{t})$,  $\epsilon_{\ul{x}}~(\epsilon_{\ul{y}})$ is a completely antisymmetric tensor. The equations \eref{fdissa2} have to be satisfied for any variation of $\ul{x}$ and $\ul{y}$ and any $j_1$ and are in general not sufficient to ensure the closeness of the hierarchy. However they are sufficient in case of general quadratic Markovian fermionic noise determined by the coupling operator basis $L_{j,k}\in\{ c_jc_k,c_j^\dag c_k^\dag,c_j^\dag c_k\}$. In this case the equations \eref{fdissa2} simplify to 
\begin{eqnarray}\nonumber
\!\!\!\!\!\!\!\!\!\!\!\!\!\!\!\!\!\!\!\!\!\!\!\!\!\!\!\!\!\!\!\!\!\!\!\!0=& 4(G^{(2,0;2,0)}_{j_1,j_2;t_1,t_2}-\bar{G}^{(0,2;0,2)}_{j_1,j_2;t_1,t_2})+ G^{(1,1;1,1)}_{j_1,j_2;t_1,t_2}-G^{(1,1;1,1)}_{j_1,t_1;j_2,t_2}-\bar{G}^{(1,1;1,1)}_{j_2,j_1;t_2,t_1}+\bar{G}^{(1,1;1,1)}_{t_1,j_1;t_2,j_2}, \\ \nonumber
\!\!\!\!\!\!\!\!\!\!\!\!\!\!\!\!\!\!\!\!\!\!\!\!\!\!\!\!\!\!\!\!\!\!\!\!0= & -2(G^{(1,1;2,0)}_{j_1,k_1;t_1,t_2}+G^{(1,1;2,0)}_{j_1,t_1;k_1,t_2}+G^{(1,1;2,0)}_{j_1,t_2;k_1,t_1})+2(-\bar{G}^{(1,1;0,2)}_{k_1,j_1;t_1,t_2}+G^{(1,1;2,0)}_{t_1,j_1;k_1,t_2}+G^{(1,1;2,0)}_{t_2,j_1;t_1,k_1}),\\ \nonumber
\!\!\!\!\!\!\!\!\!\!\!\!\!\!\!\!\!\!\!\!\!\!\!\!\!\!\!\!\!\!\!\!\!\!\!\!&+2G^{(1,1;0,2)}_{j_1,k_1;t_2,t_1}+2\bar{G}^{(1,1;2,0)}_{k_1,j_1;t_2,t_1}, \\ \nonumber
\!\!\!\!\!\!\!\!\!\!\!\!\!\!\!\!\!\!\!\!\!\!\!\!\!\!\!\!\!\!\!\!\!\!\!\!0= & G^{(2,0;0,2)}_{j_1,j_2;t_1,t_2}+G^{(2,0;0,2)}_{j_1,t_1;t_2,j_2}+G^{(2,0;0,2)}_{j_1,t_2;j_2,t_1}-\bar{G}^{(0,2;2,0)}_{j_1,j_2;t_1,t_2}-\bar{G}^{(0,2;2,0)}_{j_1,t_1;t_2,j_2}-\bar{G}^{(0,2;2,0)}_{j_1,t_2;j_2,t_1},
\end{eqnarray}
and have to be satisfied for all different values of $j_1,j_2,k_1,t_1,t_2$. 

The "higher order" requirements can be obtain in a similar manner, by writing the vanishing conditions for unwanted terms in the dissipators. 
\subsection{Mixed dissipator -- Example}
In this appendix we show an example of a mixed dissipator satisfying the third mixed closeness condition. We consider a dissipative evolution determined by the  dissipator \eref{eq:general_D} with the coupling operator basis $L_j\in\{c,\,c^\dag,\,a,\,a^\dag \}$. Since our purpose is only to give an example of a Lindblad dissipator satisfying the mixed closeness conditions we restrict ourselves to the following rate matrix
\begin{eqnarray}
{\bf G}=\left(
\begin{array}{cccc}
G_{11}&0&G_{13}&0\\
0&G_{22}&0&G_{24}\\
\bar{G}_{13}&0&G_{33}&0\\
0&\bar{G}_{24}&0&G_{44}
\end{array}
\right),
\end{eqnarray}
which for $|G_{24}|^2<|G_{22}G_{44}|$ and $|G_{13}|^2<|G_{11}G_{33}|$ defines a valid dissipator. Rewriting the dissipator in terms of the fermionic and bosonic creation and annihilation maps it is not difficult to see that the mixed closeness conditions are satisfied if $G_{11}=G_{22}$ and $\bar{G}_{13}=G_{24}$. For simplicity we write $G_{33}=G_{44}+2\delta$. In this case the dissipator may be written as
\begin{eqnarray}
\mch{D}_{\rm mixed}=&G_{11}(\hat{f}_{1}^+\hat{f}_1-\hat{f}_2^+\hat{f}_2)-\delta(\hat{b}_1^+\hat{b}_1+\hat{b}^+_2\hat{b}_2)+G_{44}\hat{b}_1^+\hat{b}_2^+\\ \nonumber 
&+\sqrt{2}(G_{13}\hat{b}^+_1\hat{f}^+_2+\bar{G}_{13}\hat{b}^+_2\hat{f}^+_1)\mch{P},
\end{eqnarray}
and satisfies the third mixed closeness condition. For the system to be stable $\delta$ should be positive, in other words the bosonic mode should be more dissipated than incoherently pumped into the system.

\section{Exact solution of the dissipative Fermi-Hubbard model}
\label{Sol_Fermi_Hub}
\begin{figure} [!!h]
\subfiguretopcaptrue
\centering{\subfigure[Density plot]{\includegraphics[width=0.4\textwidth]{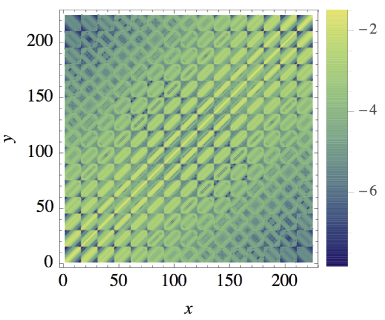}}~~~~\subfigure[Correlations with one edge.]{\includegraphics[width=0.53\textwidth]{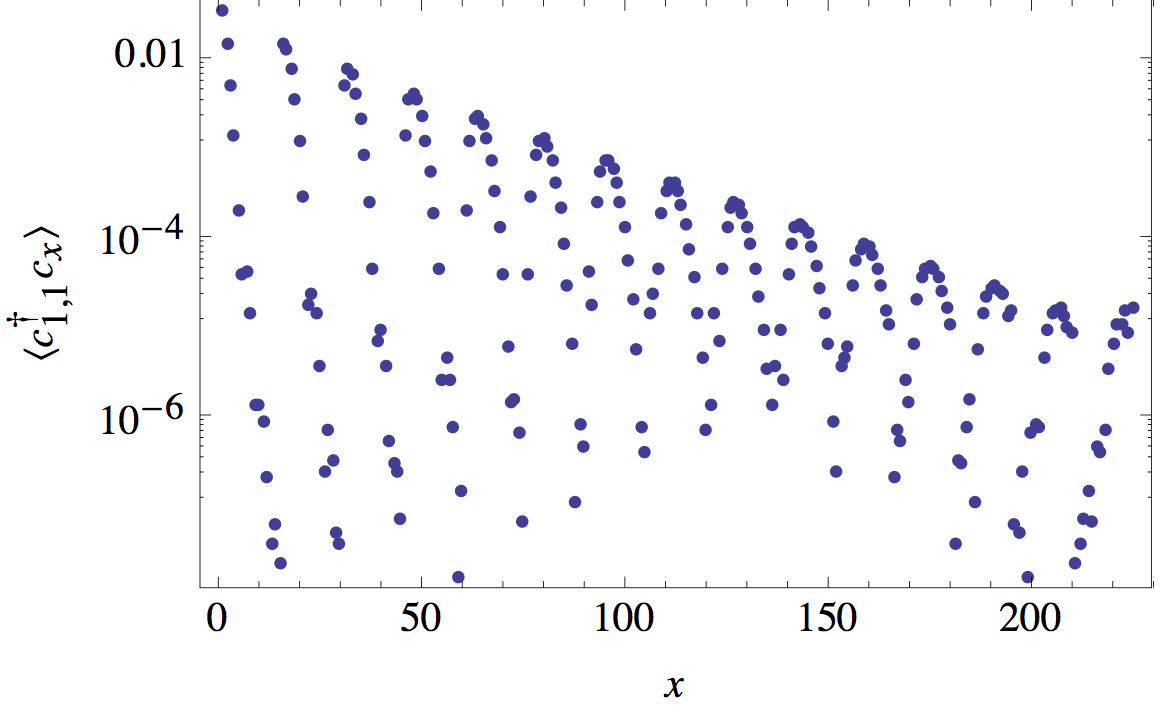} }}
\caption{Two-point correlation matrix in a 2D Fermi-Hubbard model with homogeneous nearest-neighbour interaction. a) We show a density plot of the non-vanishing correlations $\ave{c_{i,j,\sigma}^\dag c_{i',j',\sigma}}$, where $x=i+(j-1)n$, $y=i'+(j'-1)n$, and $n=15$ is number of sites along one edge. The color scale is logarithmic. b) We show one slice of the density plot, namely all non-vanishing correlations with one edge.}
\label{fig:Correlations2D}
\end{figure}

In this appendix we provide the details on the dissipative Fermi-Hubbard Liouvillian described in Section \ref{sec:Hubbard} satisfying the fermionic closeness relations. After some calculation we find that a dissipator of the form \eref{eq:general_D} defined with the rate matrix
\begin{eqnarray}
\label{fermi_G}
\!\!\!\!\!\!\!\!\!\!\!\!\!\!\!\!\!\!\!\!\!\!\!\!\!\!\!\!\!\!\!\!\!{\bf G}=\left( \begin{array}{cccccccc}
G_{11} &  0 & -G_{33}+\ii U & 0&0&0&0&0\\
0 & G_{22} & 0 & -G_{44}+\ii U&0&0 &0&0\\ 
-G_{33}-\ii U & 0 & G_{33} & 0 &0&0&0&0\\ 
0 & -G_{44}-\ii U & 0 & G_{44}&0&0&0&0\\ 
0&0&0&0&G_{33}&0&0&0\\0&0&0&0&0&G_{44}&0&0\\
0&0&0&0&0&0&G_{55}&0\\
0&0&0&0&0&0&0&G_{66}
\end{array}
\right)
\end{eqnarray}
and the dissipator basis $L_{\ul{j},\ul{k}}\in\{c_{1},c_2, c^\dag_{2}c_2 c_1,c^\dag_{1}c_1 c_2, c^\dag_{1}c_2^\dag c_2,c^\dag_{2}c_1^\dag c_1,c_1^\dag,c_2^\dag\}$ exactly cancels the closeness violating terms of the interacting part of the Fermi-Hubbard Hamiltonian \eref{Hub_int}. Further, we observe that the eigenvalues of the rate matrix \eref{fermi_G} , namely $\{G_{11}+G_{33}\pm\sqrt{(G_{11}-G_{33})^2+4(G_{33}^2+U^2)}), \,G_{22}+G_{44}\pm\sqrt{(G_{22}-G_{44})^2+4(G_{44}^2+U^2)}\,,G_{33},G_{55}\}\,$ are all positive iff  $G_{11}>G_{33}>0$ and $G_{22}>G_{44}>0$, $G_{55},G_{66}>0$, and $U^2<\min(G_{11}G_{33}-G_{33}^2,\,G_{22}G_{44}-G_{44}^2)$. For simplicity we take $G_{11}=G_{22}$, $G_{33}=G_{44}$ and $G_{55}=G_{6,6}$. In this case the Liouvillian $\mch{F}_{\rm Hub}=-\ii\,\hat{\rm ad}_{H_{\rm Hub}}+\mch{D}_{\rm Hub}$, with the Hamiltonian \eref{Hubbard_H} and the dissipator from \Sref{sec:Hubbard}, generates a completely positive evolution and may be written as
\begin{eqnarray}
\label{Hubbard_L}
\!\!\!\!\!\!\!\!\!\!\!\!\!\!\!\!\!\!\!\!\!\!\!\!\!\!\!\!\!\!\!\!\!\mch{F}_{\rm Hub}=&-\ii\sum_{\sigma=0,1}\sum_{j,k=1}^N t_{jk}\left(\hat{f}_{2j-\sigma}^+\hat{f}_{2k-\sigma}-\hat{f}_{2N+2k-\sigma}^+\hat{f}_{2N+2j-\sigma} \right)\\ \nonumber
\!\!\!\!\!\!\!\!\!\!\!\!\!\!\!\!\!\!\!\!\!\!\!\!\!\!\!\!\!\!\!\!\!&+\sum_{j=1}^N \Bigg(-\frac{1}{2}(G_{11}+ G_{33}+G_{55}) \left(\hat{f}^+_{2 j} \hat{f}_{2 j} +\hat{f}^+_{2 j-1} \hat{f}_{2j-1}+\hat{f}^+_{2 j+2 N-1} \hat{f}_{2 j+2 N-1} +\hat{f}^+_{2 j+2 N} \hat{f}_{2 j+2 N}\right)  \\ \nonumber
\!\!\!\!\!\!\!\!\!\!\!\!\!\!\!\!\!\!\!\!\!\!\!\!\!\!\!\!\!\!\!\!\!&+(G_{11}-G_{33}-G_{55}) \left(\hat{f}^+_{2 j+2 N} \hat{f}^+_{2 j}+\hat{f}^+_{2 j+2 N-1} \hat{f}^+_{2 j-1}\right) \\ \nonumber
\!\!\!\!\!\!\!\!\!\!\!\!\!\!\!\!\!\!\!\!\!\!\!\!\!\!\!\!\!\!\!\!\!& -(2G_{33}+\ii U) \hat{f}^+_{2 j}\hat{f}^+_{2 j-1}\hat{f}_{2 j}\hat{f}_{2 j-1}-(2G_{33}-\ii U)\hat{f}^+_{2 j+2 N}\hat{f}^+_{2 j+2 N-1}\hat{f}_{2 j+2 N}\hat{f}_{2 j+2 N-1} \\ \nonumber
\!\!\!\!\!\!\!\!\!\!\!\!\!\!\!\!\!\!\!\!\!\!\!\!\!\!\!\!\!\!\!\!\!& -G_{33}\Big(\hat{f}^+_{2 j+2 N-1} \hat{f}^+_{2 j-1} \hat{f}_{2 j+2 N-1} \hat{f}_{2 j-1}+  \hat{f}^+_{2 j+2 N-1} \hat{f}^+_{2 j-1} \hat{f}_{2 j+2 N} \hat{f}_{2 j} \\ \nonumber 
\!\!\!\!\!\!\!\!\!\!\!\!\!\!\!\!\!\!\!\!\!\!\!\!\!\!\!\!\!\!\!\!\!&~~~~~~~~~+ \hat{f}^+_{2 j+2 N} \hat{f}^+_{2 j} \hat{f}_{2 j+2 N-1} \hat{f}_{2 j-1} + \hat{f}^+_{2 j+2 N} \hat{f}^+_{2 j} \hat{f}_{2 j+2 N} \hat{f}_{2 j} \Big)
\\ \nonumber
\!\!\!\!\!\!\!\!\!\!\!\!\!\!\!\!\!\!\!\!\!\!\!\!\!\!\!\!\!\!\!\!\!& +\ii\,U\Big(\hat{f}^+_{2 j+2 N}\hat{f}^+_{2 j+2 N-1}\hat{f}^+_{2 j-1}\hat{f}_{2 j+2 N}-\hat{f}^+_{2 j+2 N}\hat{f}^+_{2 j+2 N-1}\hat{f}^+_{2 j}\hat{f}_{2 j+2 N-1}\\ \nonumber 
\!\!\!\!\!\!\!\!\!\!\!\!\!\!\!\!\!\!\!\!\!\!\!\!\!\!\!\!\!\!\!\!\!&~~~~~~~~~- \hat{f}^+_{2 j+2 N}\hat{f}^+_{2 j}\hat{f}^+_{2 j-1}\hat{f}_{2 j-1}+ \hat{f}^+_{2 j+2 N-1}\hat{f}^+_{2 j}\hat{f}^+_{2 j-1}\hat{f}_{2 j}\Big)\\ \nonumber
\!\!\!\!\!\!\!\!\!\!\!\!\!\!\!\!\!\!\!\!\!\!\!\!\!\!\!\!\!\!\!\!\!&-2G_{33}\Gamma \hat{f}^+_{2 j+2 N}\hat{f}^+_{2 j+2 N-1}\hat{f}^+_{2 j}\hat{f}^+_{2 j-1}+\\ \nonumber
\!\!\!\!\!\!\!\!\!\!\!\!\!\!\!\!\!\!\!\!\!\!\!\!\!\!\!\!\!\!\!\!\!&+ G_{33} \left(\hat{f}^+_{2 (j+n)} \hat{f}^+_{2 j} \hat{f}^+_{2 j-1} \hat{f}_{2 (j+n)} \hat{f}_{2
   j} \hat{f}_{2 j-1}+\hat{f}^+_{2 (j+n)} \hat{f}^+_{2 j+2 n-1} \hat{f}^+_{2 j} \hat{f}_{2 (j+n)} \hat{f}_{2 j+2
   n-1} \hat{f}_{2 j}\right.\\ \nonumber
\!\!\!\!\!\!\!\!\!\!\!\!\!\!\!\!\!\!\!\!\!\!\!\!\!\!\!\!\!\!\!\!\!&~~~~~~~~~\left.+   \hat{f}^+_{2 (j+n)} \hat{f}^+_{2 j+2 n-1} \hat{f}^+_{2 j-1} \hat{f}_{2 (j+n)} \hat{f}_{2 j+2
   n-1} \hat{f}_{2 j-1}+\hat{f}^+_{2 j+2 n-1} \hat{f}^+_{2 j} \hat{f}^+_{2 j-1} \hat{f}_{2 j+2 n-1} \hat{f}_{2
   j} \hat{f}_{2 j-1}\right)
\Bigg).
\end{eqnarray}
This form is suitable for the efficient solutions of \eref{diff_Z_fermi}. We calculate the steady-state two-point correlations in one, two, and  three dimensions. We assume nearest neighbour hopping. In one dimension all two-point correlations vanish except for the occupations which are given by $\ave{c_{j,\sigma}^\dag c_{j,\sigma}}=\frac{G_{55}}{G_{11}-G_{33}+G_{55}}$ and are independent of the length of the chain and disorder in the hopping. In two dimensions the two-point correlations $\ave{c_{j,\sigma}^\dag c_{k,\sigma}}$ do not vanish and decay exponentially with the distance from the diagonal  $|j-k|$ (see \Fref{fig:Correlations2D}); all other two-point steady state correlations still vanish. Similar indications of exponential decay of correlations are observed in three dimensions, although we also observe small correlation revivals at the edges of the chain (see main text \Fref{fig:Correlations}).

\end{document}